\documentclass[%
groupedaddress,
preprint,
showpacs,
 amsmath,amssymb,
 aps,
pra,
]{revtex4-1}

\newcommand\beq{\begin{equation}}
\newcommand\eeq{\end{equation}}

\usepackage{graphicx}% Include figure files
\usepackage{dcolumn}% Align table columns on decimal point
\usepackage{bm}% bold math
\usepackage{xcolor}
\usepackage{soul}
\begin{document}

%\preprint{APS/123-QED}

\title{Aperiodic-Order-Induced Enhancement of Weak Nonlocality in Multilayered Dielectric Metamaterials}

\author{Marino Coppolaro}
\author{Giuseppe Castaldi}
\author{Vincenzo Galdi}
\email{vgaldi@unisannio.it}
\affiliation{Fields \& Waves Lab, Department of Engineering, University of Sannio, I-82100 Benevento, Italy}

\date{\today}% It is always \today, today,

%%%%%%%%%%%%%%%%%%%% Created: 		09/04/2018
%%%%%%%%%%%%%%%%%%%% Last revised: 	30/10/2018

\begin{abstract}
Recent studies on fully dielectric multilayered metamaterials have shown that the {\em negligibly small} nonlocal effects (spatial dispersion) typically observed in the limit of {\em deeply subwavelength} layers may be significantly enhanced by peculiar boundary effects occurring in certain critical parameter regimes. 
These phenomena, observed so far in periodic and randomly disordered geometries, are manifested as strong differences between the exact optical response of finite-size metamaterial samples and the prediction from conventional effective-theory-medium models based on mixing formulae.
Here, with specific focus on the Thue-Morse geometry, we make a first step toward extending the studies above to the middle-ground of {\em aperiodically ordered} multilayers, lying in between perfect periodicity and disorder. We show that, also for these geometries, there exist critical parameter ranges that favor the buildup of boundary effects leading to strong enhancement of the (otherwise negligibly weak) nonlocality. However, the underlying mechanisms are fundamentally different from those observed in the periodic case, and exhibit typical footprints (e.g., fractal gaps, quasi-localized states) that are distinctive of aperiodic order. The outcomes of our study indicate that aperiodic order plays a key role in the buildup of the aforementioned boundary effects, and may also find potential applications to optical sensors, absorbers and lasers. 
\end{abstract}

%\pacs{}% PACS, the Physics and Astronomy
\maketitle

%%%%%%%%%%%%%%%%%%%%%%%%%%%%%%%%%%%%%%%%%%%%%%%%%%%%%%%%%%%%%%
\section{Introduction}
%%%%%%%%%%%%%%%%%%%%%%%%%%%%%%%%%%%%%%%%%%%%%%%%%%%%%%%%%%%%%%

One key feature that distinguishes optical ``metamaterials'' \cite{Capolino:2009vr,Cai:2010om,Urbas:2016ro} from other artificial materials such as photonic crystals \cite{Joannopoulos:2008pc} is the possibility to describe their macroscopic response in terms of {\em effective parameters} (e.g., permittivity and permeability), along the lines of what is conventionally done with natural materials. From the mathematical viewpoint, rigorous implementations of this modeling process, typically referred to as ``homogenization'', rely on first-principle concepts such as field averaging \cite{Smith:2006ho}. From the experimental viewpoint, such effective parameters can be retrieved via suitable measurements of the scattering matrix \cite{Smith:2005ep,Arslanagic:2013ar}.

The basic, intuitive rationale underlying homogenization is that, as long as the electrical sizes of the material inclusions are {\em very small} on the wavelength scale, and their interactions are {\em weak}, the fast field fluctuations inside the metamaterial are averaged out, and an electromagnetic wave effectively ``sees'' a continuum whose constitutive properties are dictated by mixing formulae \cite{Sihvola:1999em} which essentially depend on the inclusions' material properties, shapes, orientations and proportions, but not on their sizes and spatial order. To give an example that is especially relevant for the present study, in a multilayered metamaterial composed by stacking two types of deeply subwavelength material layers (with distinct constitutive properties and thicknesses, labeled, e.g., with ``$a$'' and ``$b$''), the effective parameters should depend on the filling fractions (i.e., proportions of the $a$- and $b$-type constituents in the mixture) but not on the specific order and/or arrangement of the layers, so that configurations associated with sequences such as $abababab$, $babababa$ and $abbaabba$ should all be effectively equivalent, and should all differ from, e.g., $aaabaaba$ \cite{Sihvola:1999em}. 

The inherent limitations and range of applicability of the simple ``effective-medium theory'' (EMT) above are well known, and more complex extensions have been developed to capture the spatial-dispersion (nonlocal) effects \cite{Landau:1960eo,Agranovich:2013co} which may become non-negligible, e.g., in the presence of electrically thick and/or metallic inclusions (see, e.g., Refs. \cite{Silveirinha:2007mh,Alu:2011fp,Ciattoni:2015nh} and
\cite{Elser:2007ne,Chebykin:2011ne,Chebykin:2012ne,Chern:2013sd} for general and multilayer-specific approaches, respectively). For instance, in multilayered metamaterials, the presence of metallic layers (albeit deeply subwavelength) may induce strong nonlocal effects, due to the coherent interactions of surface-plasmon-polaritons \cite{Maier:2007pf} propagating at the metal-dielectric interfaces, which can manifest as the appearance of additional extraordinary waves \cite{Orlov:2011eo} not predicted by the EMT.

Much less expectable and counterintuitive is the ``breakdown'' of the EMT in periodic multilayered metamaterials with {\em fully dielectric}, {\em deeply subwavelength} layers, which was recently predicted on theoretical grounds by Herzig Sheinfux {\em et al.} \cite{Sheinfux:2014sm}, and experimentally observed by Zhukovsky {\em et al.} \cite{Zhukovsky:2015ed}. Basically, it was shown that, under specific illumination settings, the optical response (transmittance or reflectance) of finite-thickness samples may exhibit substantial differences from the EMT prediction, accompanied by an ultrasensitivity to the spatial arrangement, size and termination of the layers. As also elucidated in follow-up studies \cite{Andryieuski:2015ae,Popov:2016oa,Lei:2017rt,Maurel:2018so,Castaldi:2018be}, these phenomena are not manifested in the {\em bulk} (infinite-medium) response, and can be interpreted as {\em boundary effects} stemming from the peculiar, interface-dominated phase-accumulation mechanism in the multilayer, which may strongly enhance the (otherwise negligibly weak) nonlocality. These effects can be captured by suitable nonlocal extensions \cite{Popov:2016oa,Castaldi:2018be}.
Related theoretical \cite{Sheinfux:2016cr} and experimental \cite{Sheinfux:2017oo} studies in similar parameter regimes, but characterized by {\em random} spatial disorder, have evidenced the possibility to attain Anderson localization, likewise in stark contrast with the EMT prediction, and once again with ultrasensitivity to changes of features on a deeply subwavelength scale. These results have sparked considerable interest, both in terms of implications for the homogenization theory, and potential applications to extreme optical sensing and switching.

Against the background above, this study explores the possibility to observe similar effects in {\em aperiodically ordered} geometries, i.e., the vast middle ground separating perfect periodicity and random disorder. Originally inspired by the concept of ``quasicrystals'' in solid-state physics \cite{Shechtman:1984mp,Levine:1984qa}, aperiodic order has become increasingly relevant in many fields of science and technology \cite{Macia:2006tr} and, in particular, in optics and photonics \cite{DalNegro:2011da} (see also a related perspective in a recent roadmap on optical metamaterials \cite[Sec. 3]{Urbas:2016ro}). As a representative geometry, we consider the Thue-Morse (ThM) sequence \cite{Queffelec:2010sd}, which has been extensively studied in the past in connection with photonic crystals \cite{Liu:1997po,Qiu:2003rt,DalNegro:2004pb,Jiang:2005pb,Hiltunen:2008lt,Grigoriev:2010bm,Grigoriev:2010bs,Hsueh:2011fo} and metallo-dielectric multilayers \cite{Savoia:2013on}, but has never been explored in the fully dielectric, deeply subwavelength regime of interest here.

Accordingly, the paper is structured as follows. In Sec. \ref{Sec:Statement}, we introduce the problem geometry and its formulation. In Sec. \ref{Sec:Model}, we describe the modeling tools and related mathematical formalism based on the trace and antitrace maps \cite{Wang:2000ta}. In Sec. \ref{Sec:Results}, we discuss some representative results. Finally, in Sec. \ref{Sec:Conclusions}, we draw some conclusions and point to future work.

%%%%%%%%%%%%%%%%%%%%%%%%%%%%%%%%%%%%%%%%%%%%%%%%%%%%%%%%%%%%%%
\section{Problem Statement}
%%%%%%%%%%%%%%%%%%%%%%%%%%%%%%%%%%%%%%%%%%%%%%%%%%%%%%%%%%%%%%
\label{Sec:Statement}
%-------------------------------------------------------------
\subsection{Geometry}
\begin{figure}
	\centering
	\includegraphics[width=12cm]{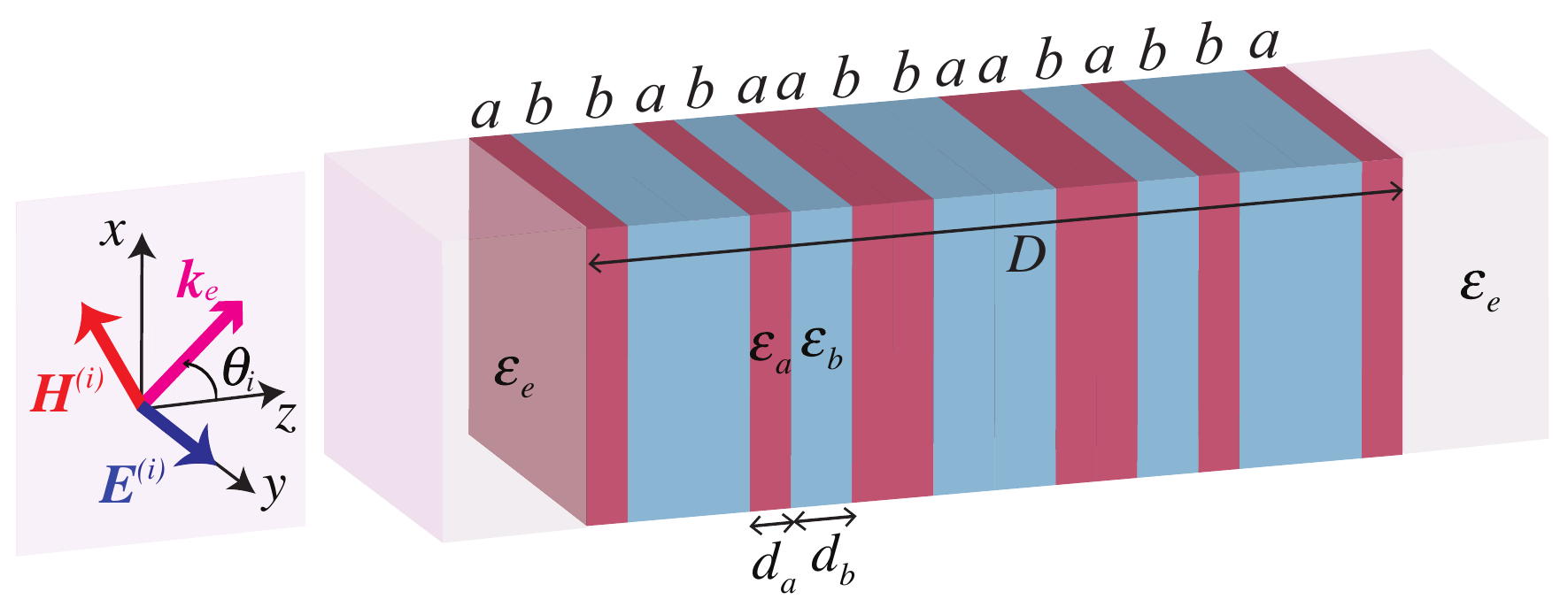}
	\caption{Problem geometry illustrating the ThM multilayered metamaterial and illumination conditions (details in the text). }
	\label{Figure1}
\end{figure}

%-------------------------------------------------------------
The problem geometry is schematized in Fig. \ref{Figure1}. We consider a multilayer composed of two types of dielectric layers (labeled as ``$a$'' and ``$b$''), with relative permittivity $\varepsilon_a$ and $\varepsilon_b$, and thickness $d_a$ and $d_b$, stacked along the $z$-direction, and of infinite extent in the $x-y$ plane. The layers are arranged aperiodically according to the ThM sequence, generated by assuming the symbol $a$ as an initiator, and iteratively applying the substitution rules \cite{Queffelec:2010sd}
\beq
a\rightarrow ab,\quad b\rightarrow ba.
\eeq
The first iterations are therefore $a$, $ab$, $abba$, $abbabaab$, and so on, with the generic $n$th stage of growth containing $N=2^n$ layers. As general, well-known traits of this sequence, we recall that at any iteration $n\ge1$: {\em i)} the frequency of occurrence of $a$- and $b$-type symbols is identical (and hence exactly the same as for periodic sequences), {\em ii)}  each half of the sequence corresponds to the ``flipped'' version of the other half, and {\em iii)} sequences containing more than two consecutive symbols (e.g., $aaa$ or $bbb$) are not possible \cite{Queffelec:2010sd}.

In what follows, we consider a generic multilayer at stage of growth $n$, with total thickness $D=2^{n-1} d$ (with $d=d_a+d_b$ denoting the thickness of an $ab$-type bilayer), embedded in a homogeneous dielectric medium with relative permittivity $\varepsilon_e$. For instance, the case depicted in Fig. \ref{Figure1} corresponds to the stage of growth $n=4$ (i.e., $N=16$ layers).

%-------------------------------------------------------------
\subsection{Formulation}
%-------------------------------------------------------------

For illumination, we assume a time-harmonic plane wave, with suppressed $\exp\left(-i\omega t\right)$ time dependence and transverse-electric (TE) polarization ($y$-directed electric field), impinging with an angle $\theta_i$ from the $z$-axis, viz.,
\beq
E_y^{(i)}=E_0\exp\left[i \left(k_x x+
k_{ze}z
\right)\right],
\label{eq:Ei}
\eeq
where $E_0$ denotes a real-valued amplitude, and
\beq
k_x=k\sqrt{\varepsilon_e}\sin\theta_i,\quad k_{ze}=k\sqrt{\varepsilon_e}\cos\theta_i
\label{eq:ke}
\eeq
are the transverse (conserved) and longitudinal components, respectively, of the wavevector ${\bf k}_e$ (see Fig. \ref{Figure1}). In Eq. (\ref{eq:ke}), $k=\omega/c$ is the vacuum wavenumber, with $c$ denoting the corresponding wavespeed.
As previously mentioned, we assume to operate in the {\em deeply subwavelength} regime, i.e., $d_{a}, d_b \ll \lambda$, with $\lambda=2\pi/k$ denoting the vacuum wavelength. Under these conditions, the optical response of the multilayer is generally well captured by an EMT model in terms of a homogeneous, uniaxially anisotropic slab characterized by a relative permittivity tensor whose parallel ($\parallel$, i.e., $x-y$) and orthogonal ($\perp$, i.e., $z$) components are given by simple Maxwell-Garnett-type mixing formulae \cite{Sihvola:1999em}
\begin{subequations}
\begin{eqnarray}
{\bar \varepsilon}_{\parallel}&=&f_a \varepsilon_a+f_b\varepsilon_b,
\label{eq:epspar}\\
{\bar \varepsilon}_{\perp}&=&\left(f_a \varepsilon_a^{-1}+f_b\varepsilon_b^{-1}\right)^{-1},
\end{eqnarray}
\label{eq:EMT}
\end{subequations}
with $f_a=d_a/d$ and $f_b=d_b/d=1-f_a$ denoting the filling fractions pertaining to $a$- and $b$-type constituents, respectively, and the overbar utilized throughout the paper to indicate EMT-based quantities. First, we observe that the mixing formulae in Eqs. (\ref{eq:EMT}) are exactly identical with those pertaining to a conventional periodic multilayer (repetitions of $ab$-type bilayers). This should not be surprising, as we have previously recalled that, just like the periodic ones, ThM sequences exhibit the same distribution of $a$- and $b$-type symbols, and that EMT models are sensitive to proportions, rather than spatial arrangement. It is also worth pointing out that, in view of the assumed TE polarization, only the parallel component in Eq. (\ref{eq:epspar}) is actually relevant to our study.

Previous studies on ThM-based optical structures have focused on photonic crystals (i.e., moderately thick layers) \cite{Liu:1997po,Qiu:2003rt,DalNegro:2004pb,Jiang:2005pb,Hiltunen:2008lt,Grigoriev:2010bm,Grigoriev:2010bs,Hsueh:2011fo} and hyperbolic metamaterials (i.e., deeply subwavelength metallic and dielectric layers) \cite{Savoia:2013on}, which exhibit a wealth of interesting effects such as bandgaps, resonant transmission, localization and field enhancement, omnidirectional reflection, fractal edge-states, multistability, additional extraordinary waves.

Conversely, in what follows, we deal with ThM-based metamaterials featuring fully dielectric, deeply subwavelength layers, and study the possible buildup of boundary effects leading to strong enhancement the (otherwise negligibly weak) nonlocality. To this aim, we
systematically compare the exact optical response of structures at various stages of growth, and under different illumination conditions, with the corresponding EMT-based predictions, in order to identify critical parameter regimes where nonlocality may be strongly enhanced. Moreover, to single out behaviors that are genuinely induced by the underlying aperiodic order, we also consider the comparison with the well-established periodic-multilayer case \cite{Sheinfux:2014sm,Zhukovsky:2015ed,Andryieuski:2015ae,Popov:2016oa,Lei:2017rt,Maurel:2018so,Castaldi:2018be} which, as observed above, shares the same EMT model.

%%%%%%%%%%%%%%%%%%%%%%%%%%%%%%%%%%%%%%%%%%%%%%%%%%%%%%%%%%%%%%
\section{Modeling Tools and Formalism}
%%%%%%%%%%%%%%%%%%%%%%%%%%%%%%%%%%%%%%%%%%%%%%%%%%%%%%%%%%%%%%
\label{Sec:Model}

%-------------------------------------------------------------
\subsection{Transfer-Matrix Model}
%-------------------------------------------------------------
The optical response of the ThM multilayered metamaterial in Fig. \ref{Figure1} can be rigorously calculated by means of the well-established transfer-matrix method \cite[Chap. 1]{Born:1999un}. Basically, the transverse field components at the two interfaces of a generic  $a$- or $b$-type layer can be related via
\beq
\left[
\begin{array}{cc}
	E_y^{(L)}\\
	iZ_e H_x^{(L)}
\end{array}
\right]={\underline {\underline {\cal M}}}_{\nu}\cdot \left[
\begin{array}{cc}
	E_y^{(R)}\\
	iZ_e H_x^{(R)}
\end{array}
\right],
\label{eq:TM}
\eeq
where the superscripts $(L)$ and $(R)$ denote the left and right interfaces, respectively, 
\beq
Z_e=\frac{\omega \mu_0}{k_{ze}},
\eeq
represents the TE wave impedance in the exterior medium (with $\mu_0$ denoting the vacuum magnetic permeability), and  
\beq
	{\underline {\underline {\cal M}}}_{\nu}=
	\left[
	\begin{array}{cc}
	\cos\left(k_{z\nu}d_{\nu}\right) & \displaystyle{\frac{k_{ze}}{k_{z\nu}}}\sin\left(k_{z\nu}d_{\nu}\right)\\
	-\displaystyle{\frac{k_{z\nu}}{k_{ze}}}\sin\left(k_{z\nu}d_{\nu}\right)	 & \cos\left(k_{z\nu}d_{\nu}\right)
	\end{array}
	\right],
	\label{eq:Matnu}
\eeq
is a $2\times 2$, unimodular, adimensional matrix, where $\nu=a$ or $b$, and
\beq
k_{z\nu}=\sqrt{k^2\varepsilon_{\nu}-k_x^2}=k\sqrt{\varepsilon_{\nu}-\varepsilon_e\sin^2\theta_i}
\eeq
denote the longitudinal wavenumbers in the two corresponding media \cite[Chap. 1]{Born:1999un}. The representation above can readily be iterated to deal with multiple cascaded layers, via chain product of the single-layer transfer matrices \cite[Chap. 1]{Born:1999un}. 
Accordingly, we can relate the fields at the input ($z=0$) and output ($z=D$) interfaces of a ThM multilayer at stage of growth $n$ as
\begin{subequations}
\begin{eqnarray}
\left.
\left[
\begin{array}{cc}
	E_y\\
	iZ_e H_x
\end{array}
\right]\right|_{z=0}
&=&{\underline {\underline {\cal M}}}^{(n)}\cdot \left.
\left[
\begin{array}{cc}
	E_y\\
	iZ_e H_x
\end{array}
\right]\right|_{z=D},
\label{eq:Etot}\\
{\underline {\underline {\cal M}}}^{(n)}&=&\prod_{j=1}^{N=2^n}{\underline {\underline {\cal M}}}_{\nu\left(j\right)}=
\left[
\begin{array}{cc}
m^{(n)}_{11} & m^{(n)}_{12}\\
m^{(n)}_{21} & m^{(n)}_{22}
\end{array}
\right],
\label{eq:TMtot}
\end{eqnarray}
\label{eq:MML}
\end{subequations}
with $\nu\left(j\right)=a$ or $b$, according to the $jth$-symbol in the ThM sequence. 

Based on the model in Eqs. (\ref{eq:MML}), for a given incident field, we can rigorously calculate the reflection and transmission coefficients, as well as the field distribution inside the multilayer.

%-------------------------------------------------------------
\subsection{Trace and Antitrace Maps}
%-------------------------------------------------------------
As already pointed out in our previous study dealing with the periodic case \cite{Castaldi:2018be}, some key observables in the optical response of a generic multilayer can be calculated without the need to actually perform the chain matrix product in Eq. (\ref{eq:TMtot}), which, for a large number of layers, may become both computationally intensive and prone to numerical-error propagation. For instance, by defining the transmission coefficient
\beq
\tau_n=
\frac{\left. E_y^{(t)}\right|_{z=D}}{\left. E_y^{(i)}\right|_{z=0}},
\label{eq:taun}
\eeq
with the superscript $(t)$ tagging the transmitted field, we obtain from Eq. (\ref{eq:Etot}) (see Appendix \ref{Sec:AppA} for details)
\beq
\tau_n=\frac{2}{m^{(n)}_{11}+m^{(n)}_{22}+i\left[m^{(n)}_{21}-m^{(n)}_{12}\right]}=\frac{2}{\mbox{Tr}\left[{\underline {\underline {\cal M}}}^{(n)}\right]+
i\mbox{ATr}\left[{\underline {\underline {\cal M}}}^{(n)}\right]}, 
\label{eq:tau}
\eeq
where $\mbox{Tr}\left[\cdot\right]$ and $\mbox{ATr}\left[\cdot\right]$ denote the conventional matrix {\em trace} and {\em antitrace} operators, respectively \cite{Lang:1987la}. Quite remarkably, similar to the periodic multilayer case \cite{Castaldi:2018be}, also for the ThM geometry of interest here it is possible to compute these quantities iteratively via simple polynomial maps. More specifically, by letting
\beq
\chi_n\equiv\mbox{Tr}\left[{\underline {\underline {\cal M}}}^{(n)}\right], \quad
\upsilon_n\equiv\mbox{Atr}\left[{\underline {\underline {\cal M}}}^{(n)}\right],\quad
{\tilde \upsilon_n}\equiv\mbox{Atr}\left[{\underline {\underline {\cal \tilde M}}}^{(n)}\right]
\label{eq:tat},
\eeq 
with the tilde denoting a complementary configuration featuring a ThM sequence initiated with a $b$-type (instead of $a$-type) symbol, it can be shown \cite{Wang:2000ta,Grigoriev:2010bs,Savoia:2013on} that the evolution with respect to the stage of growth $n$ is ruled by the following intertwined maps
\begin{subequations}
\begin{eqnarray}
\chi_{n+2}&=&\chi_n^2\left(\chi_{n+1}-2\right)+2,
\label{eq:TMchi}\\
\upsilon_{n+1}&=&\chi_{n-1}\left[
\left(
\chi_n-1
\right)\upsilon_{n-1}+{\tilde \upsilon}_{n-1}
\right],\\
{\tilde \upsilon}_{n+1}&=&\chi_{n-1}\left[
\left(
\chi_n-1
\right){\tilde \upsilon}_{n-1}+\upsilon_{n-1}
\right],~~~n\ge1,
\end{eqnarray}
\label{eq:TaTmaps}
\end{subequations}
where $\chi_0=\mbox{Tr}\left({\underline {\underline {\cal M}}}_a\right)$, $\upsilon_0=\mbox{Atr}\left({\underline {\underline {\cal M}}}_a\right)$ and ${\tilde \upsilon}_0=\mbox{Atr}\left({\underline {\underline {\cal M}}}_b\right)$. 

Though directly related to the evolution of the transmission coefficient [see Eq. (\mbox{\ref{eq:tau})}], trace and antitrace are not physically meaningful quantities, and hence cannot be used \emph{per se} in order to ascertain the enhancement of nonlocality. Nevertheless, possible departures of the maps in Eqs. (\mbox{\ref{eq:TaTmaps}}) from the corresponding EMT (local) predictions effectively quantify the degree of nonlocality. Within this framework, 
for the periodic multilayer case \cite{Castaldi:2018be}, we showed that the buildup of boundary effects leading to the enhancement of nonlocality could be effectively interpreted
and parameterized in closed-form in terms of error propagation in the trace and antitrace maps. In the ThM case of interest here, 
the trace and antitrace maps in Eqs. (\ref{eq:TaTmaps}) cannot be solved analytically in closed form. Nevertheless, 
the interpretation of the boundary effects in terms of error propagation still holds. 
It is worth stressing that the derivation of the trace and antitrace maps is \emph{exact}, and therefore the computation of the transmission coefficient via
Eqs. (\mbox{\ref{eq:tau}}) and (\mbox{\ref{eq:TaTmaps}}) is fully equivalent to that arising from the chain matrix product in Eqs. (\mbox{\ref{eq:MML}}). In addition, the trace-antitrace-map scheme is also computationally more effective and robust with respect to roundoff errors, as well as more insightful. 

%%%%%%%%%%%%%%%%%%%%%%%%%%%%%%%%%%%%%%%%%%%%%%%%%%%%%%%%%%%%%%
\section{Representative Results}
%%%%%%%%%%%%%%%%%%%%%%%%%%%%%%%%%%%%%%%%%%%%%%%%%%%%%%%%%%%%%%
\label{Sec:Results}

%-------------------------------------------------------------
\subsection{Parameters and Observables}
%-------------------------------------------------------------
To facilitate comparison with previous studies on periodic and random structures, we consider the same material parameters as in Refs. \mbox{\cite{Sheinfux:2014sm,Castaldi:2018be,Lei:2017rt,Sheinfux:2016cr}}, for the layers ($\varepsilon_a=1$, $\varepsilon_b=5$, possibly with some small loss-gain perturbations) and exterior medium ($\varepsilon_e=4$), with identical filling fractions $f_a=f_b=0.5$ (i.e., $d_a=d_b=d/2$), which correspond to an EMT relative permittivity ${\bar \varepsilon_{\parallel}}=3$. Likewise, we mainly focus on parameter configurations where the field is propagating in the higher-permittivity layers and in the effective medium, and evanescent in the lower-permittivity ones. This corresponds to an angular incidence range
\beq
\theta_{ac}\equiv
\arcsin\left(
\sqrt{\frac{\varepsilon_a}{\varepsilon_e}}
\right)<
\theta_i
\lesssim \arcsin\left(
\sqrt{\frac{{\bar \varepsilon}_{\parallel}}{\varepsilon_e}}
\right)\equiv{\bar \theta}_c.
\label{eq:theta_crit}
\eeq
As for the electrical thickness, we explore the range $0.04<d/\lambda<0.1$, which guarantees that the layers remain deeply subwavelength.
Our parametric studies below consider ThM multilayers at various stages of growth $n$, which correspond to $N=2^n$ layers. 

For the lossless scenarios, besides the trace $\chi_n$ and antitrace $\upsilon_n$, we consider as the main physical observables the transmittance
\beq
T_n=\left|\tau_n\right|^2=\frac{4}{\left|\chi_n+i\upsilon_n\right|^2},
\label{eq:Tn}
\eeq
and the electric field (magnitude) distribution in the multilayer [computed by means of the transfer-matrix chain in Eqs. (\mbox{\ref{eq:MML}})]. For scenarios featuring optical losses or gain, we also consider the reflectance
\beq
R_n=\left|\rho_n\right|^2,
\label{eq:Rn}
\eeq
computed from the reflection coefficient (see Appendix \ref{Sec:AppA} for details)
\beq
\rho_n=
\left.
\frac{E_y^{(r)}}{E_y^{(i)}}\right|_{z=0}=\tau_n\left[m^{(n)}_{11}-im^{(n)}_{12}\right]-1,
\label{eq:rhon}
\eeq
with the superscript $(r)$ tagging the reflected field. From Eq. (\ref{eq:rhon}), we observe that, unlike the transmittance, the reflectance does not depend solely on the trace and antitrace. We stress that, in principle, it is possible to derive evolution maps [formally similar to those in Eqs. (\ref{eq:TaTmaps})] for {\em any} of the transfer-matrix elements \cite{Wang:2000ta}. These, however, are not reported here for brevity.

For lossy scenarios, we also compute the absorbance, which follows directly from power conservation:
\beq
A_n=1-T_n-R_n.
\label{eq:An}
\eeq

To ascertain the possible enhancement of nonlocal effects, and the role played by aperiodic order, we also study the two reference configurations considered in Ref. \cite{Castaldi:2018be}, namely, a homogeneous slab with relative permittivity ${\bar \varepsilon}_{\parallel}$ given by the EMT model in Eq. (\ref{eq:epspar}) and thickness $D$, and a periodic multilayer with same type and total number of layers (and hence thickness $D$). In both cases, the observables above can be computed analytically. For the homogeneous EMT slab, they can be computed from a single transfer matrix as in Eq. (\ref{eq:Matnu}) (by assuming $\varepsilon_{\nu}={\bar \varepsilon}_{\parallel}$, $d_{\nu}=D$), while for the periodic case they readily follow from the closed-form solutions of the trace and antitrace maps \cite[Eqs. (14)]{Castaldi:2018be}.

%-------------------------------------------------------------
\subsection{Lossless Case}
\begin{figure}
	\centering
	\includegraphics[width=16cm]{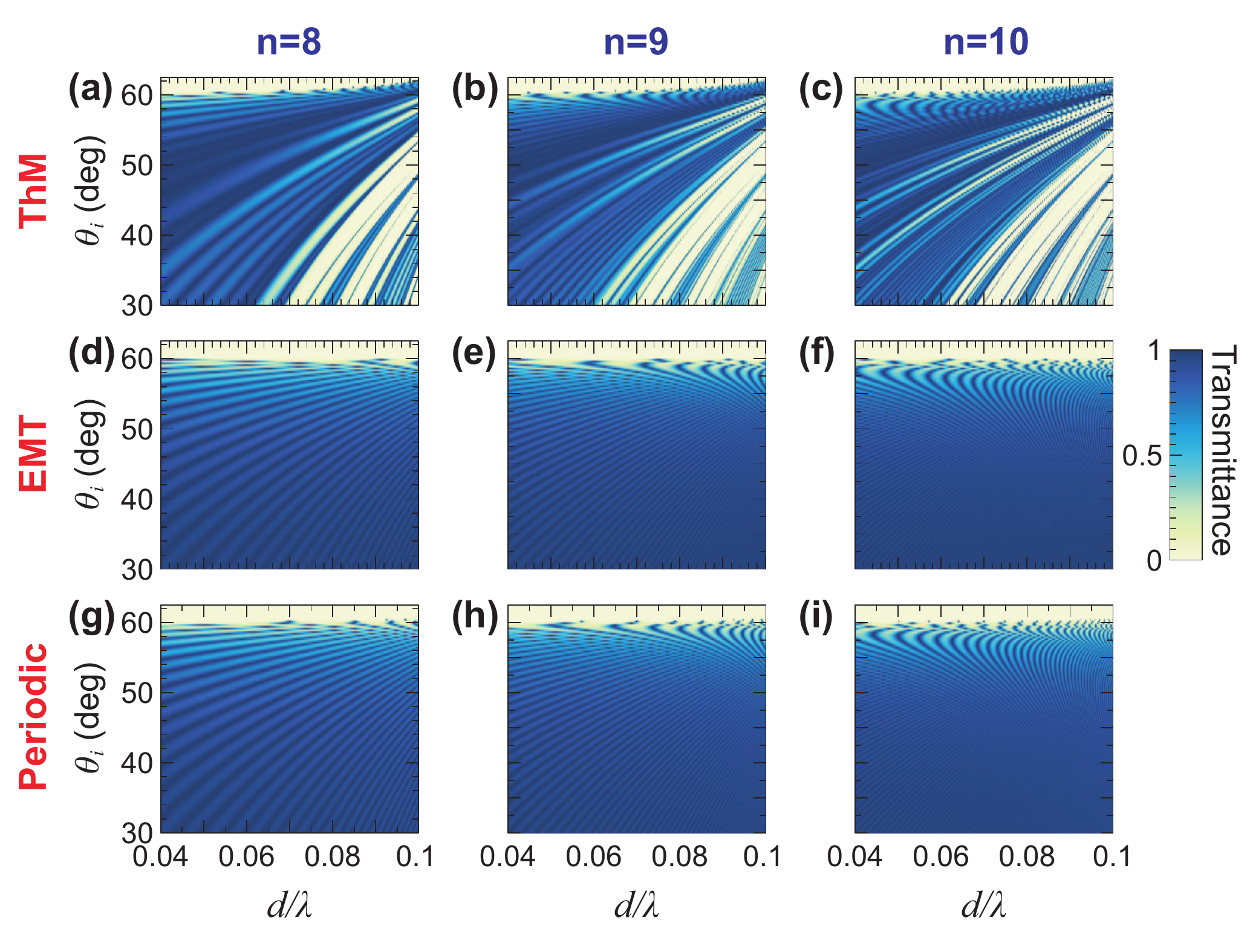}
	\caption{(a), (b), (c) Transmittance [see Eq. (\ref{eq:Tn})] responses pertaining to ThM multilayered metamaterials at stages of growth $n=8$ ($N=256$ layers), $n=9$ ($N=512$ layers), and $n=10$ ($N=1024$ layers), respectively, for $\varepsilon_a=1$, $\varepsilon_b=5$, $f_a=f_b=0.5$ (i.e., $d_a=d_b=d/2$), and $\varepsilon_e=4$, as a function of the electrical thickness $d/\lambda$ and incidence angle $\theta_i$. (d), (e), (f) Same as above, but EMT predictions (${\bar\varepsilon}_{\parallel}=3$). (g), (h), (i) Same as above, but for periodic arrangements.}
	\label{Figure2}
\end{figure}

%-------------------------------------------------------------
We start considering the lossless scenario ($\varepsilon_a=1$, $\varepsilon_b=5$). Figure \ref{Figure2} compares the transmittance responses of three representative ThM configurations at various stages of growth ($n=8,9,10$) with the corresponding EMT and periodic benchmarks, as a function of the electrical thickness $d/\lambda$ of the $ab$-type bilayer (henceforth, simply referred to as ``electrical thickness'' for compactness) and the incidence direction $\theta_i$. This latter, according to  Eq. (\ref{eq:theta_crit}), varies within the range $\theta_{ac}=30^o<\theta_i\lesssim {\bar \theta}_c=60^o$. At a qualitative glance, we observe a generally good agreement between the EMT [Figs. \ref{Figure2}(d)--(f)] and periodic [Figs. \ref{Figure2}(g)--(i)] responses, which exhibit the expectable small-to-moderate Fabry-P\'erot-type oscillations of the transmittance, with the possible exception of the region nearby the critical angle ${\bar \theta}_c=60^o$, where the field undergoes a transition from propagating to evanescent in the effective medium. Conversely, the ThM responses [Figs. \ref{Figure2}(a)--(c)] exhibit a markedly different behavior also far away from the critical angle, with much more pronounced (bandgap-like) oscillations.
In what follows, we examine in more detail two distinctive mechanisms underlying these strong departures.  

%=============================================================
\subsubsection{Near-Critical Incidence}
%=============================================================
In the periodic-multilayer case \cite{Sheinfux:2014sm,Zhukovsky:2015ed,Andryieuski:2015ae,Popov:2016oa,Lei:2017rt,Maurel:2018so,Castaldi:2018be}, significant differences between the exact optical response and its EMT prediction were observed in the vicinity of the critical angle for which the field becomes evanescent in the effective medium ($\theta_i\lesssim {\bar \theta}_c$). In particular, we showed in Ref. \cite{Castaldi:2018be} that the trace and antitrace maps pertaining to the multilayer and a homogeneous EMT slab periodically depart according to a two-scale oscillatory law, whose maximum amplitudes may diverge asymptotically in the antitrace case (together with the slow scale) as the incidence direction approaches the critical angle ${\bar \theta}_c$.

\begin{figure}
	\centering
	\includegraphics[width=10cm]{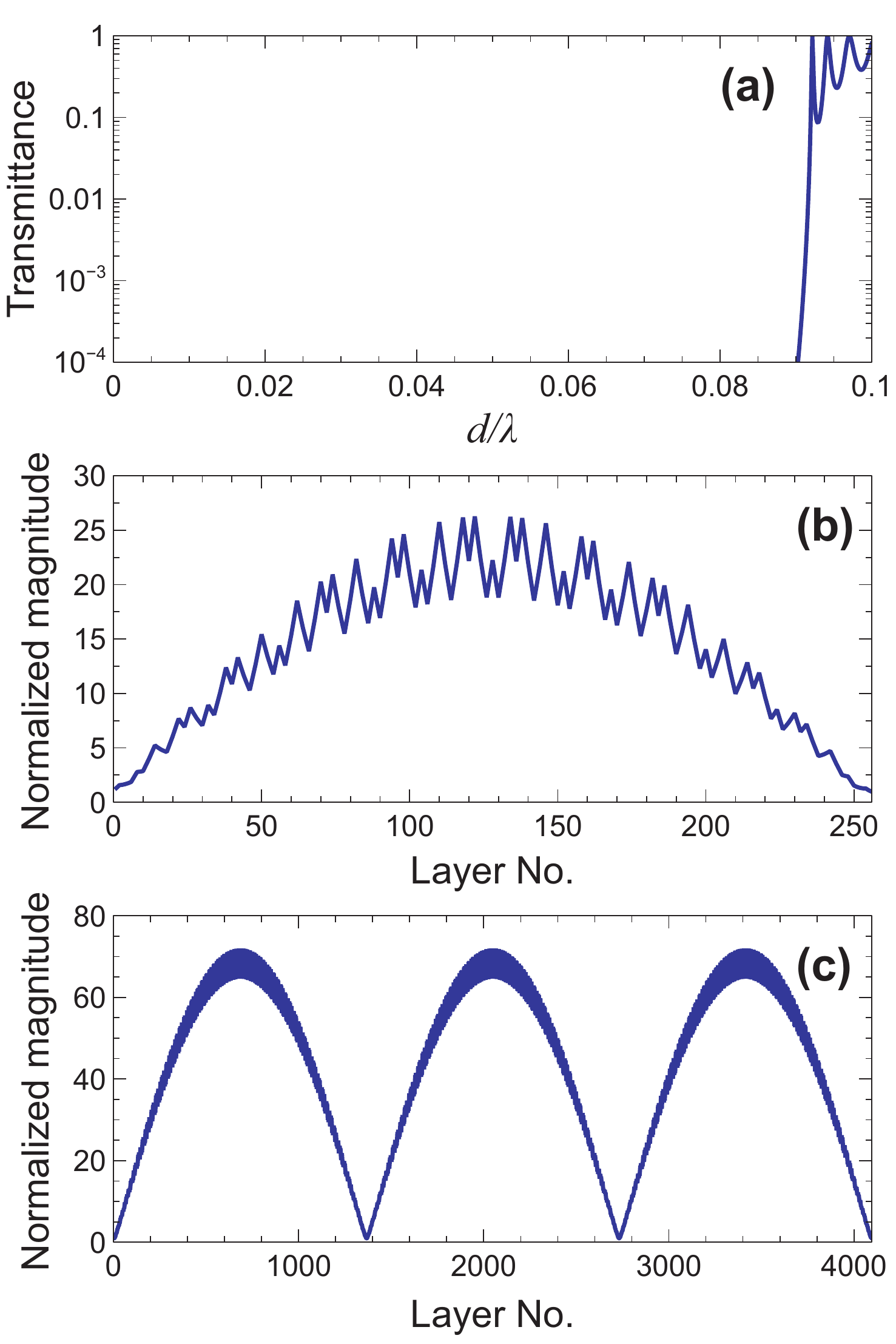}
	\caption{Parameters as in Fig. \ref{Figure2}. (a) Transmittance-cut from Fig. \ref{Figure2}(a) [ThM multilayer at stage of growth $n=8$ ($N=256$ layers)], at $\theta_i=61.85^o$. The corresponding EMT and periodic reference response (not shown), are negligibly small ($<10^{-8}$). (b) Field (magnitude) distribution inside the multilayer (normalized by the incident-field amplitude $E_0$) at $d/\lambda=0.092$. (c) Same as panel (b), but at stage of growth $n=12$ ($N=4096$ layers).}
	\label{Figure3}
\end{figure}

For the ThM case of interest here, this regime remains critical, and other interesting effects appear, which have no counterpart in the periodic scenario. Figure \ref{Figure3}(a) shows a representative transmittance cut from Fig. \ref{Figure2}(a) (ThM multilayer at stage of growth $n=8$, i.e., $N=256$ layers), at a fixed incidence angle $\theta_i=61.85^o\gtrsim {\bar \theta}_c$, for which the field is evanescent in the effective medium. As it can be observed, the transmittance is very low within most of the electrical-thickness range, but some high-transmittance resonant peaks appear for $d/\lambda\gtrsim 0.09$. Conversely, the transmittance for the corresponding EMT and periodic reference configurations remains always negligibly small ($<10^{-8}$). Associated with the high-transmittance peaks are some Fabry-P\'erot-type states, as shown for in Fig. \ref{Figure3}(b), which can exhibit strong field enhancements. Another representative example of such states is shown in Fig. \ref{Figure3}(c), for a higher stage of growth ($n=12$, i.e., $N=4096$ layers).

\begin{figure}
	\centering
	\includegraphics[width=10cm]{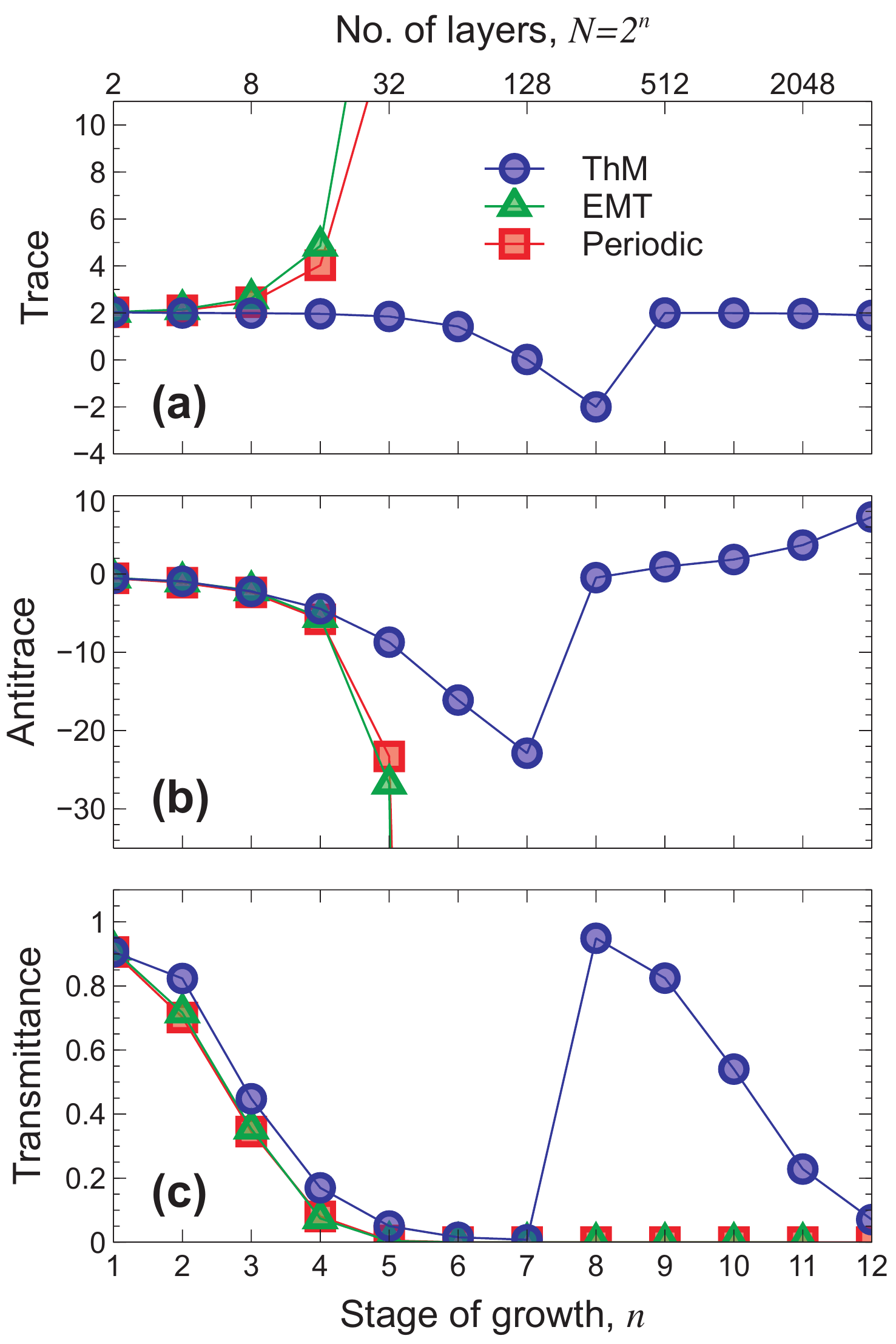}
	\caption{Parameters as in Fig. \ref{Figure2}, with $d/\lambda=0.092$ and $\theta_i=61.85^o$. (a), (b), (c) Comparisons between the trace (blue circles), antitrace (green triangles) and transmittance (red squares) evolutions pertaining to ThM, EMT and periodic configurations, respectively, as a function of the stage of growth $n$. The corresponding number of layers is also shown on the top axis. Continuous curves are guides to the eye only.}
	\label{Figure4}
\end{figure}

To gain some insight in this EMT-breakdown mechanism, which has no counterpart in the periodic case, it is instructive to look at the trace and antitrace maps. 
For fixed electrical thickness and incidence direction [corresponding to the Fabry-P\'erot-type state in Fig. \ref{Figure3}(b)], Fig. \ref{Figure4} compares the evolution of the ThM trace, antitrace and transmittance, as a function of the stage of growth $n$, with those pertaining to the EMT and periodic configurations. 
We highlight that, for all three cases, trace and antitrace start from very similar values at the initial stage of growth, namely $\chi_1=2.028$, $\upsilon_1=-0.550$ for ThM and periodic cases, and ${\bar\chi}_1=2.037$, ${\bar\upsilon}_1=-0.482$ in the EMT case, as an effect of the very weak nonlocality. These values are only slightly beyond the ``band-edge'' condition $\chi=2$ ($k_z=0$) for Bloch-type terminations \cite{Wang:2000ta}, which indicates the occurrence of a bandgap. As the structure size increases, the maps pertaining to the EMT and periodic cases exhibit similar exponentially increasing behaviors, in stark contrast with those pertaining to the ThM case, which remain bounded. Interestingly, the corresponding transmittance initially follows the rapid, monotonic decrease of the EMT and periodic cases, but then abruptly exhibits a ``revival'' at higher stages of growth ($n\ge8$).
From the mathematical viewpoint, these behaviors can be understood in terms of distinctive properties of the trace and antitrace maps. For the periodic case, it is clear from the closed form solution in Ref. \cite[Eqs. (14)]{Castaldi:2018be} that an initial condition $|\chi_1|>2$ (i.e., in a bandgap)
will inevitably lead to an exponentially increasing behavior, which is physically consistent with the evanescent character of the field. This is not true for the ThM map in Eqs. (\ref{eq:TMchi}), which can oscillate around the band-edge condition $|\chi|=2$, thereby allowing a revival of the transmittance at higher stages of growth. Although the ThM trace and antitrace maps in Eqs. (\mbox{\ref{eq:TaTmaps}}) may actually exhibit periodic orbits \mbox{\cite{Wang:2000ta}}, this is not the case for the parameter configuration in Fig. \mbox{\ref{Figure4}}, in spite of the seeming periodicity (with $n$) of ThM transmittance. For instance, no revivals are observed for values $13\le n\le20$ (not shown).

The above mechanism constitutes a first example of how negligibly weak nonlocality can be enhanced by boundary effects so as to yield strong departures of the optical response from the EMT prediction. We stress that, although these effects are still manifested in the near-critical-incidence regime $\theta_i\approx {\bar \theta}_c$, they differ fundamentally from those observed in the periodic-multilayer case 
\cite{Sheinfux:2014sm,Zhukovsky:2015ed,Andryieuski:2015ae,Popov:2016oa,Lei:2017rt,Maurel:2018so,Castaldi:2018be}, and can be genuinely attributed to the ThM aperiodic order.

%=============================================================
\subsubsection{Fractal Gaps and Quasi-Localized States}
\begin{figure}
	\centering
	\includegraphics[width=10cm]{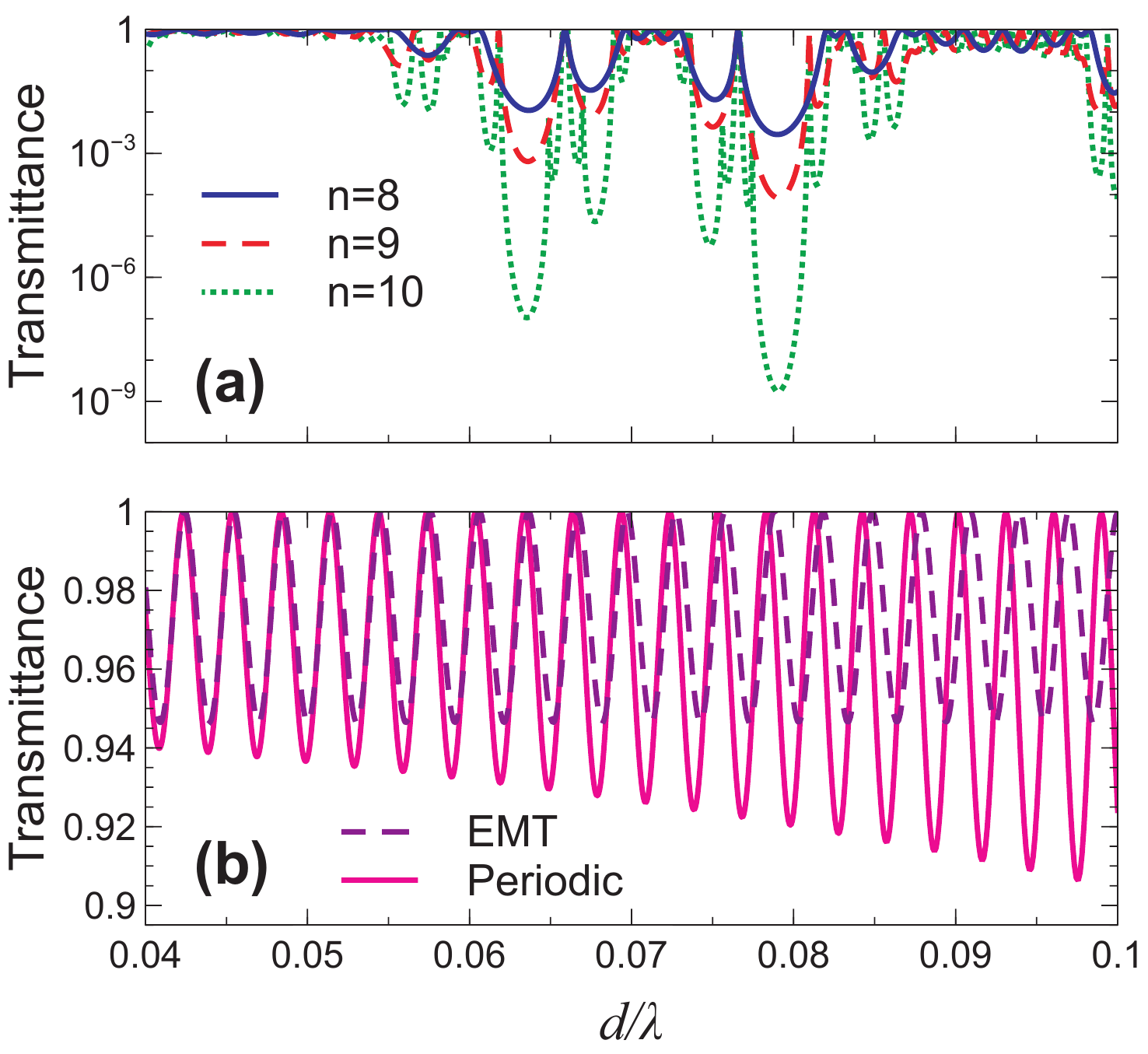}
	\caption{Parameters as in Fig. \ref{Figure2}. (a) Transmittance-cuts at $\theta_i=35.35^o$ from Fig. \ref{Figure2}(a), for ThM multilayers at stages of growth $n=8$ ($N=256$ layers; blue-solid), $n=9$ ($N=512$ layers; red-dashed), and $n=10$ ($N=1024$ layers; green-dotted). Note the logarithmic scale and very large dynamics on the vertical axis. (b), (c) Same as panel (a) but for the corresponding EMT (purple-dashed) and periodic configurations (magenta-solid) at  stage of growth $n=8$ [i.e., cuts from Figs. \ref{Figure2}(b) and \ref{Figure2}(c), respectively]. Note the linear scale and much smaller dynamics on the vertical axis.}
	\label{Figure5}
\end{figure}

%=============================================================
Away from the near-critical incidence above, there are other distinctive features that emerge in the ThM optical response. 
For a more quantitative assessment of the visual impression from Figs. \ref{Figure2}(a)--(c), Fig. \ref{Figure5}(a) shows three representative cuts at fixed incidence angle ($\theta_i=35.35^o$) and different stages of growth. For increasing size of the multilayer, we observe the formation of gaps with growing complexity that resemble fractal-type structures. This is very different from the behavior of the EMT and periodic reference responses [cf. Fig. \ref{Figure5}(b)], which are in good agreement and exhibit only small oscillations around a near-unit transmittance.

Fractal gaps are actually a well-known hallmark of ThM-based structures. Previous studies on photonic crystals \cite{Jiang:2005pb,Lei:2007pb} have explained the underlying mechanism in terms of distinctive interface correlation, and have derived the condition for a fractal gap to occur in terms of a minimal bilayer electrical thickness
\beq
\frac{\sqrt{\varepsilon_a}d_a+\sqrt{\varepsilon_b}d_b}{\lambda}=\frac{1}{3},
\label{eq:FG}
\eeq
which, for our assumed parameters, corresponds to $d/\lambda=0.206$. 
Quite remarkably, we observe similar effects at deeply subwavelength thicknesses $d/\lambda\approx0.06$, i.e., by a factor $\sim 3.5$ time smaller than the value in Eq. (\ref{eq:FG}). We also point out that our propagation regime is fundamentally different from that considered in \cite{Jiang:2005pb,Lei:2007pb}, which assumes propagating fields in both types of material layers. Instead, our assumption in Eq. (\ref{eq:theta_crit}) implies that the field is propagating in the $b$-type layers and evanescent in the $a$-type ones. As a consequence, the phase accumulation is dominated by discrete jumps at the interfaces, rather than the propagation across the layers \cite{Sheinfux:2014sm}.

\begin{figure}
	\centering
	\includegraphics[width=10cm]{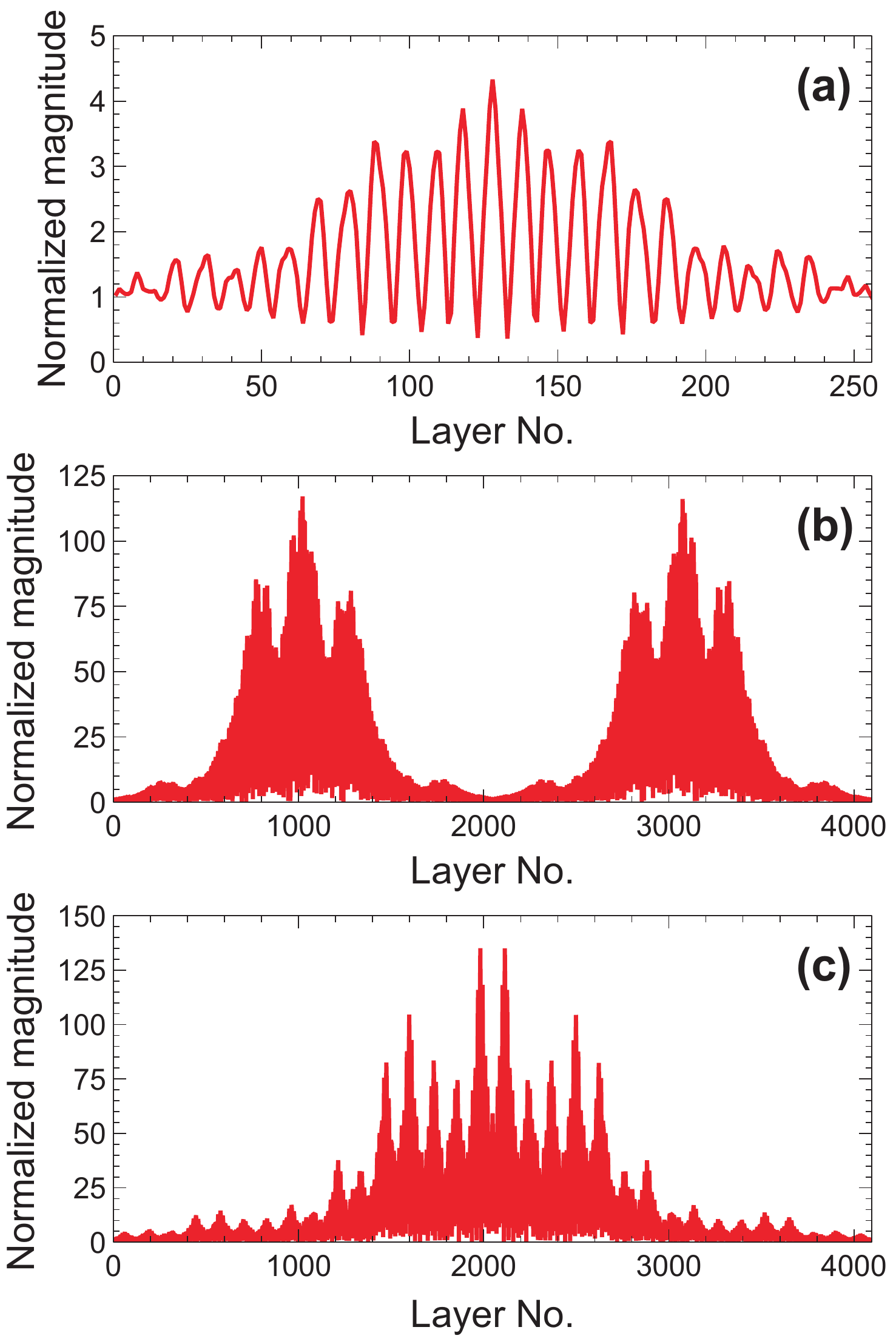}
	\caption{Parameters as in Fig. \ref{Figure2}. Field (magnitude) distributions inside the ThM multilayer (normalized by the incident-field amplitude $E_0$) for representative quasi-localized states.
		(a) $d/\lambda=0.077$, $\theta_i=35.35^o$, $n=8$ ($N=256$).  
		(b) $d/\lambda=0.062$, $\theta_i=35.35^o$, $n=12$ ($N=4096$). (c) $d/\lambda=0.085$, $\theta_i=49.6^o$, $n=12$ ($N=4096$).}
	\label{Figure6}
\end{figure}

Figure \ref{Figure6} illustrates an interesting feature that is typically associated with fractal gaps in ThM optical structures, i.e., the appearance (at the lower or upper gap-edges) of states with {\em hyperexponential} localization properties lying somewhere in between the exponential decay of localized states and the extended character of Bloch-like gap-edge states in periodic structures \cite{Jiang:2005pb,Lei:2007pb,Hiltunen:2008lt}. More specifically, Figs. \ref{Figure6}(a) and \ref{Figure6}(b) show two such states for the parameters as in Fig. \ref{Figure5}, at two representative stages of growth, whereas Fig. \ref{Figure6}(c) illustrates an example at a different incidence angle. Similar to what observed in the photonic-crystal regime \cite{Jiang:2005pb,Lei:2007pb,Hiltunen:2008lt}, for increasing stages of growth, these ``quasi-localized'' states tend to exhibit cluster-periodic distributions
with large magnitude fluctuations and strong field enhancement, and can attain very large quality factors, of potential interest for applications to optical cavities.

To give an idea, the EMT prediction for the maximum field enhancement inside the multilayer can be expressed as (see Appendix \ref{Sec:AppB} for details)
\beq
{\bar\gamma}\equiv\frac{\max\limits_{0<z<D}\left|{\bar E}_y\left(z\right)\right|}{E_0}=
\frac{k_{ze}}{{\bar k}_z}=
\frac{\sqrt{\varepsilon_e}\cos\theta_i}{\sqrt{{\bar\varepsilon}_{\parallel}-\varepsilon_e\sin^2\theta_i}},
\label{eq:enh}
\eeq
with ${\bar k}_z=\sqrt{k^2 {\bar \varepsilon}_{\parallel}-k_x^2}$ denoting the longitudinal wavenumber in the effective medium. For the parameters as in Figs. \ref{Figure6}(a) and \ref{Figure6}(b), the EMT prediction in Eq. (\ref{eq:enh}) yields a very modest ($\sim 1.26$) enhancement, in stark contrast with the actual values observed (4.3 and 116, respectively).

\begin{figure}
	\centering
	\includegraphics[width=10cm]{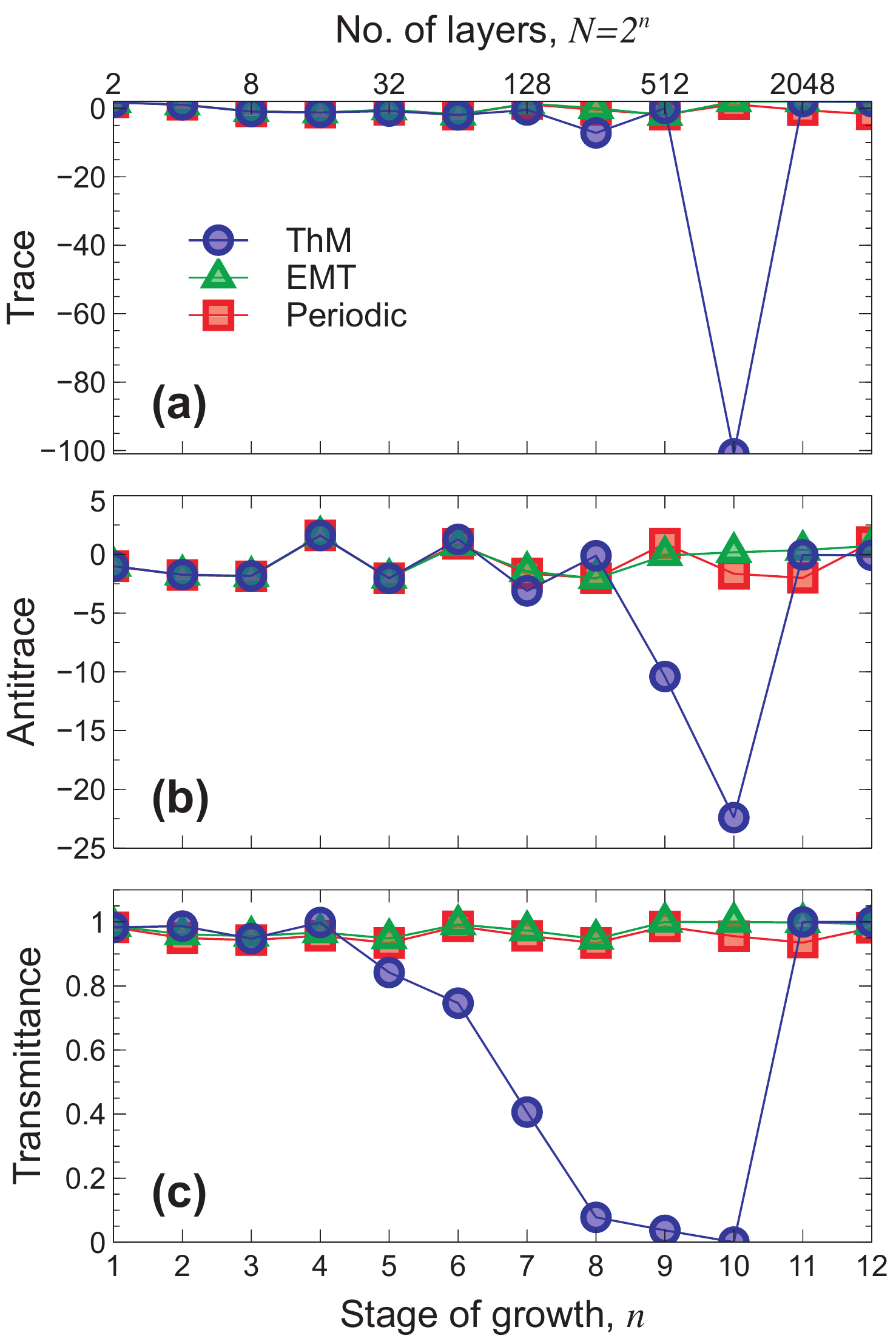}
	\caption{Parameters as in Fig. \ref{Figure6}(b). (a), (b), (c) Comparisons between the trace (blue circles), antitrace (green triangles) and transmittance (red squares) evolutions pertaining to ThM, EMT and periodic configurations, respectively, as a function of the stage of growth $n$. The corresponding number of layers is also shown on the top axis. Continuous curves are guides to the eye only.}
	\label{Figure7}
\end{figure}

It is also instructive to look at the trace and antitrace maps. Figure \ref{Figure7} compares the evolutions of trace, antitrace and transmittance for the  parameter configuration pertaining to the quasi-localized state in Fig. \ref{Figure6}(b). As it can be observed, the EMT and periodic values maintain a generally good agreement, with only moderate oscillations in the trace and antitrace and near-unit transmittance, whereas the ThM ones exhibit markedly different behaviors for intermediate stages of growth $7\le n\le10$. Qualitatively similar results can be observed in connection with other quasi-localized states.

The above results are a clear manifestation of a fundamentally different type of boundary effects, which can occur far away from the critical incidence, but are still genuinely induced by the ThM aperiodic order.

In what follows, with a view towards possible applications to sensing, absorbers and lasing, we study the effects of small losses and gain.

%-------------------------------------------------------------
\subsection{Small Losses or Gain}
%-------------------------------------------------------------
We add a small imaginary part to the permittivity of the $b$-type material, i.e., by assuming $\varepsilon_b=5+i\delta$, with $|\delta|\ll 1$; for the assumed time-harmonic convention, positive and negative values of $\delta$ correspond to optical losses and gain, respectively.

\begin{figure}
	\centering
	\includegraphics[width=10cm]{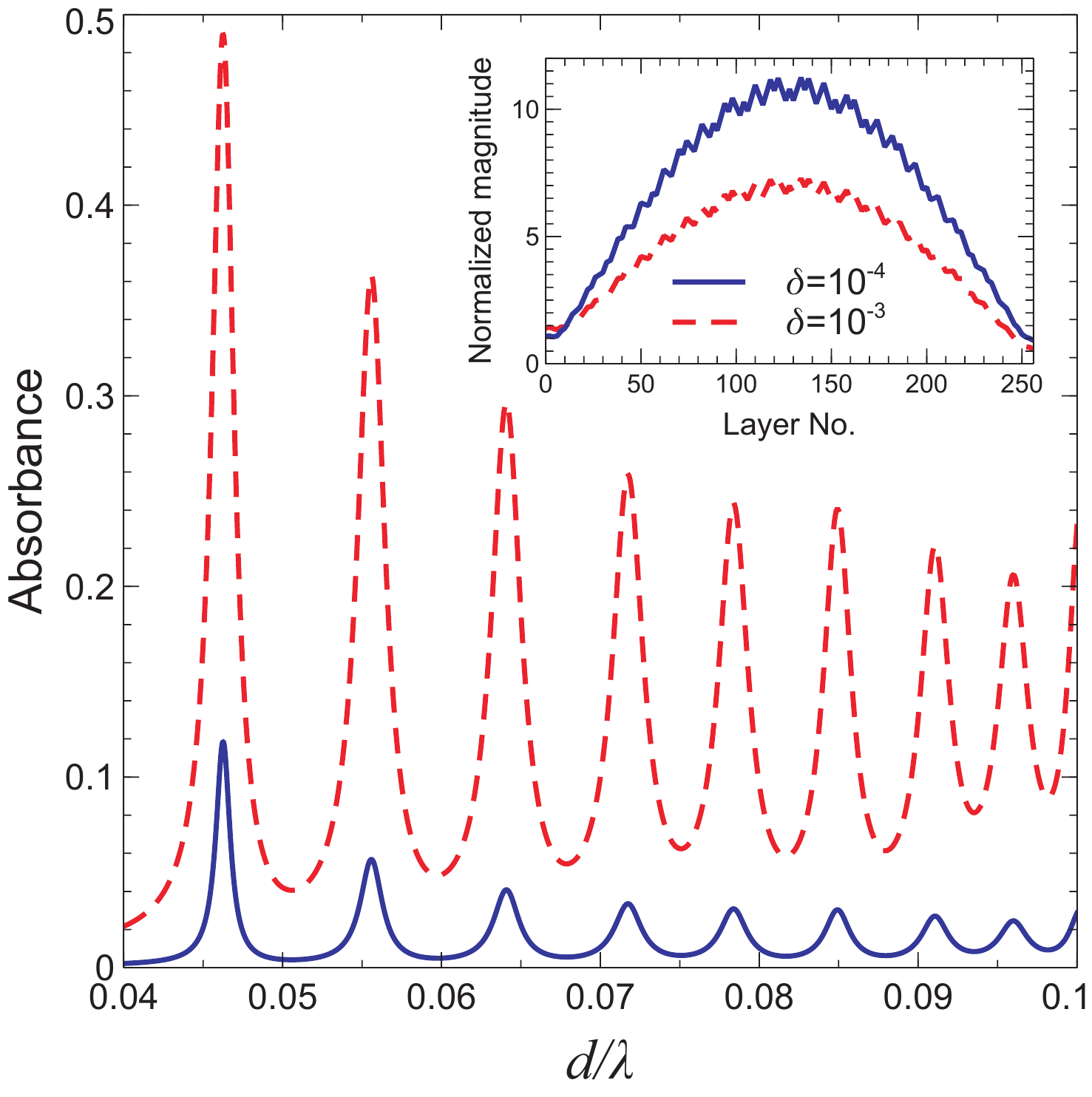}
	\caption{Parameters as in Fig. \ref{Figure2}, but with $\varepsilon_b=5+i\delta$ (with $\delta>0$, i.e., losses), $n=8$ (i.e., $N=256$ layers), and near-critical incidence $\theta_i=60.35^o$. Absorbance as a function of the electrical thickness, for $\delta=10^{-4}$ (blue-solid) and  $\delta=10^{-3}$ (red-dashed). The inset shows the normalized (magnitude) field distributions of the Fabry-P\'erot states corresponding to the peaks at $d/\lambda=0.046$.}
	\label{Figure8}
\end{figure}

Figure \ref{Figure8} shows some representative absorbance responses for the stage of growth $n=8$ ($N=256$ layers), near-critical incidence ($\theta_i=60.35^o$), and $\delta=10^{-4}$ and $10^{-3}$ (i.e., losses). As it can be observed, there are a series of resonant peaks, even for electrical thicknesses as small as $d/\lambda=0.046$, with significant values of absorbance (up to nearly 0.5). By contrast, the absorption in the EMT and periodic counterparts is negligible (on the order of $\delta$), since the field is evanescent inside the structure and gets almost completely reflected (${\bar R}_n\sim 0.999$). Also shown in the inset are the field distributions pertaining to the resonant peaks at $d/\lambda=0.046$, which display the Fabry-P\'erot-type structure already observed in the lossless case.

\begin{figure}
	\centering
	\includegraphics[width=10cm]{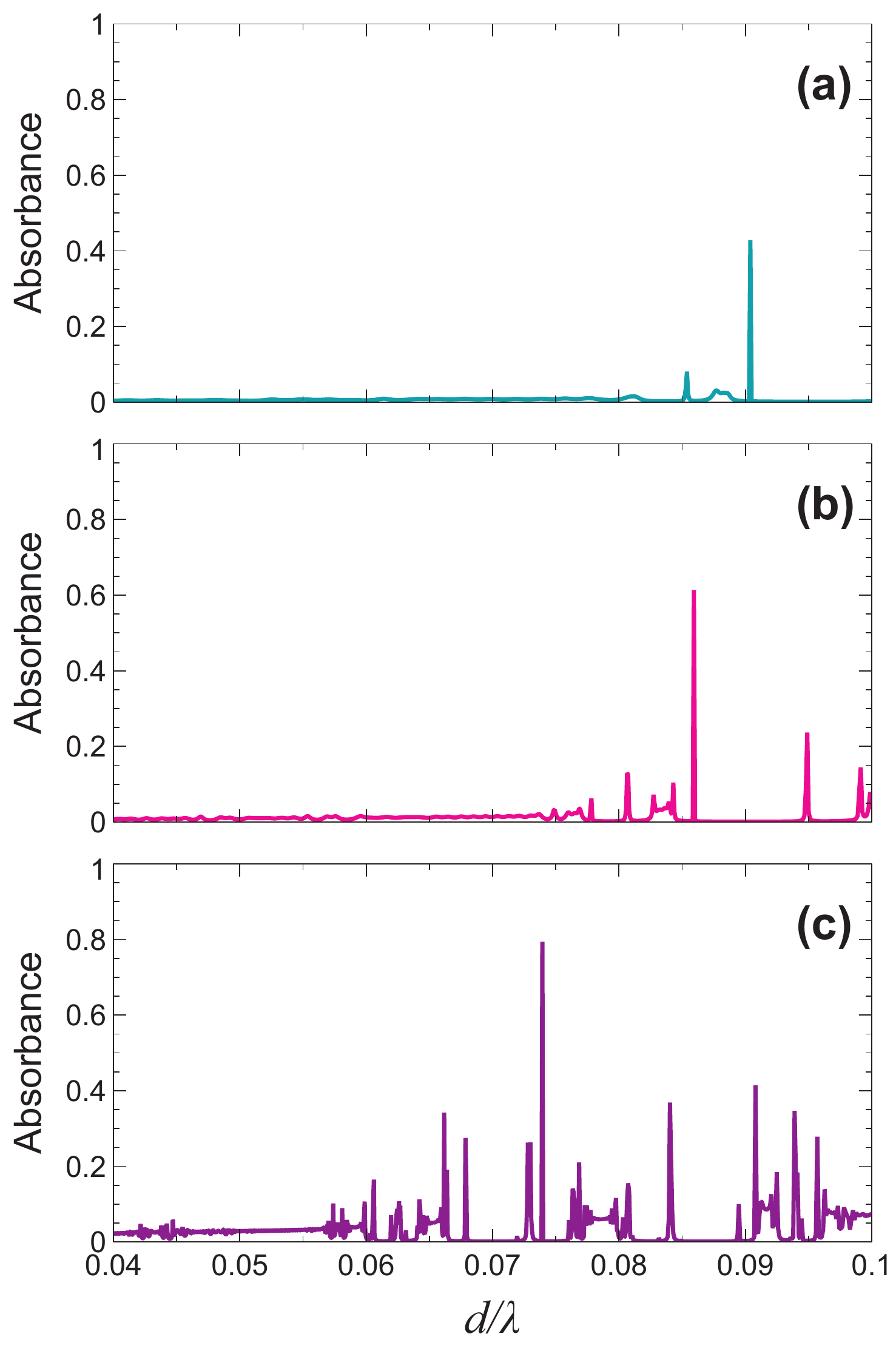}
	\caption{Parameters as in Fig. \ref{Figure2}, but with $\varepsilon_b=5+i10^{-4}$ (losses).
		Absorbance [from Eq. (\ref{eq:An})] as a function of the $ab$-type bilayer electrical thickness, for (a) $n=9$
		($N=512$ layers), $\theta_i=50.35^o$, (b) $n=10$ ($N=1024$ layers), $\theta_i=48.85^o$, (c) $n=12$ ($N=4096$ layers), $\theta_i=40.1^o$.}
	\label{Figure9}
\end{figure}

\begin{figure}
	\centering
	\includegraphics[width=10cm]{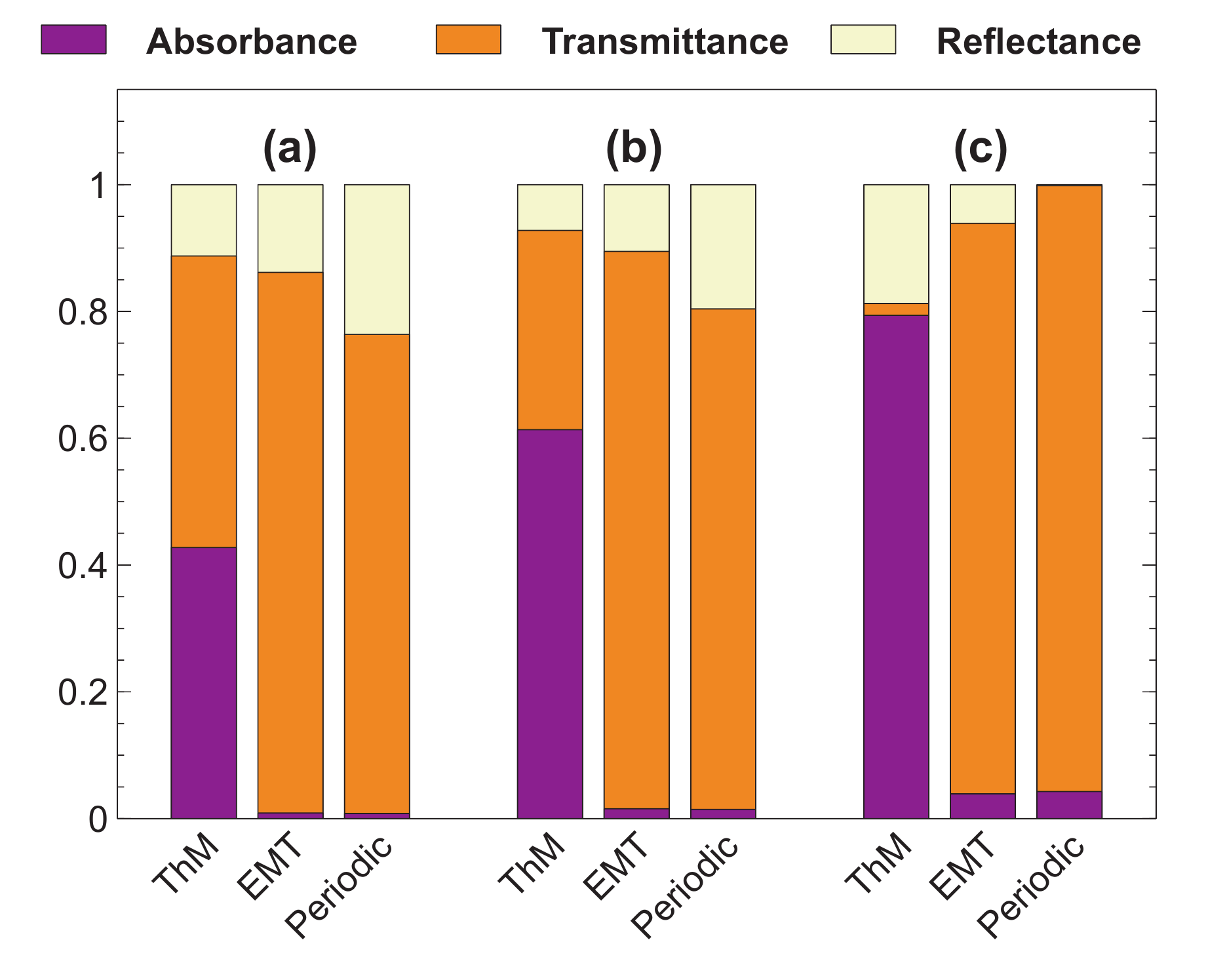}
	\caption{(a), (b), (c) Bar diagrams illustrating the absorbance, transmittance and reflectance at three representative peaks ($d/\lambda=0.090, 0.086, 0.074$, respectively) in the corresponding panels of Fig. \ref{Figure9}.}
	\label{Figure10}
\end{figure}

\begin{figure}
	\centering
	\includegraphics[width=10cm]{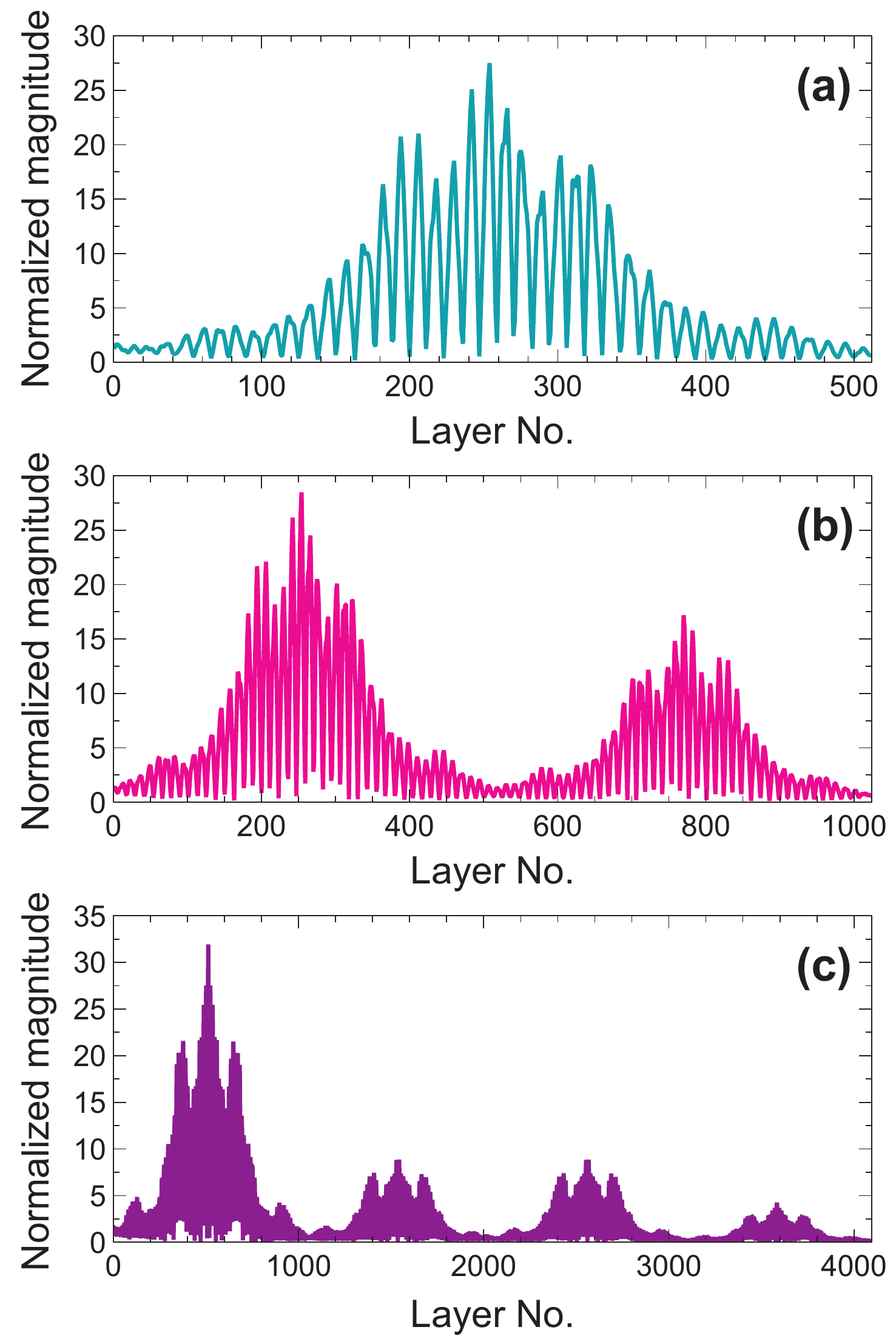}
	\caption{(a), (b), (c) Normalized (magnitude) field distributions of the quasi-localized states corresponding to the absorbance peaks in Fig. \ref{Figure10}.}
	\label{Figure11}
\end{figure}

Figure \ref{Figure9} shows some representative results in the vicinity of fractal gaps, for $\delta=10^{-4}$ and different stages of growth. Once again, several sharp peaks are observed, with absorbance as high as 0.8. For three representative peaks, the comparison with the EMT and periodic counterparts is illustrated in Fig. \ref{Figure10} in terms of bar diagrams. In all three examples, the EMT and periodic look comparable, and substantially different from the ThM counterparts. For instance, the absorbance in the ThM case is significantly higher (by a factor 20--50), and also the transmittance is quite different.
As it can be observed in Fig. \ref{Figure11}, the field distributions at the resonant peaks exhibit the quasi-localized characteristics typical of fractal-gap-edge states (cf. Fig. \ref{Figure6}).  

We highlight that the results above pertain to specific parameter values, and in general the absorbances exhibited by the three configurations (ThM, periodic, EMT) are comparable.

\begin{figure}
	\centering
	\includegraphics[width=10cm]{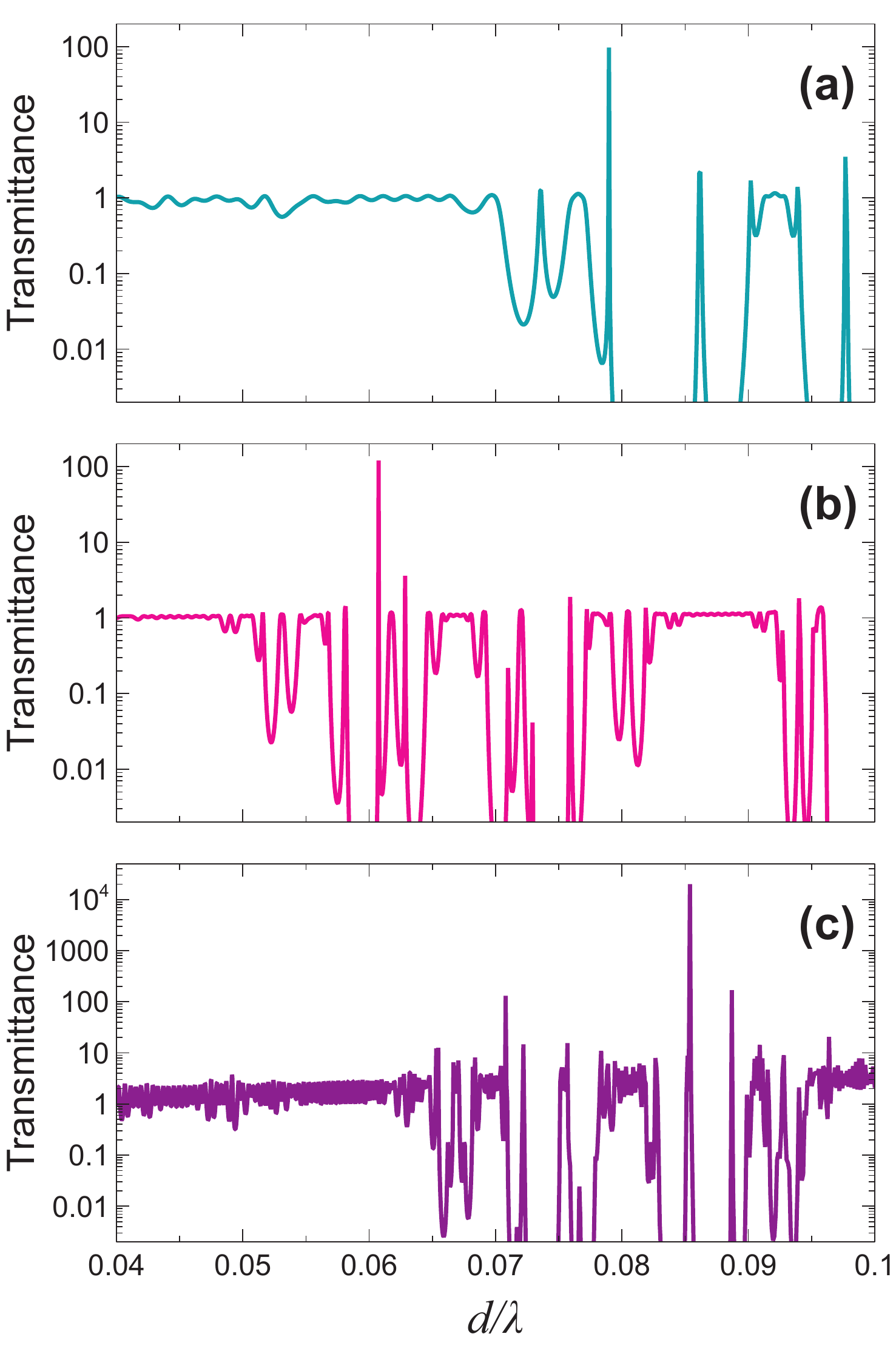}
	\caption{Parameters as in Fig. \ref{Figure2}, but with $\varepsilon_b=5-i10^{-3}$ (gain).
		Transmittance as a function of the $ab$-type bilayer electrical thickness for
		(a) $n=9$ ($N=512$ layers), $\theta_i=46.10^o$, (b) $n=10$ ($N=1024$ layers), $\theta_i=31.60^o$, (c) $n=12$ ($N=4096$ layers), $\theta_i=56.1^o$.}
	\label{Figure12}
\end{figure}

\begin{figure}
	\centering
	\includegraphics[width=10cm]{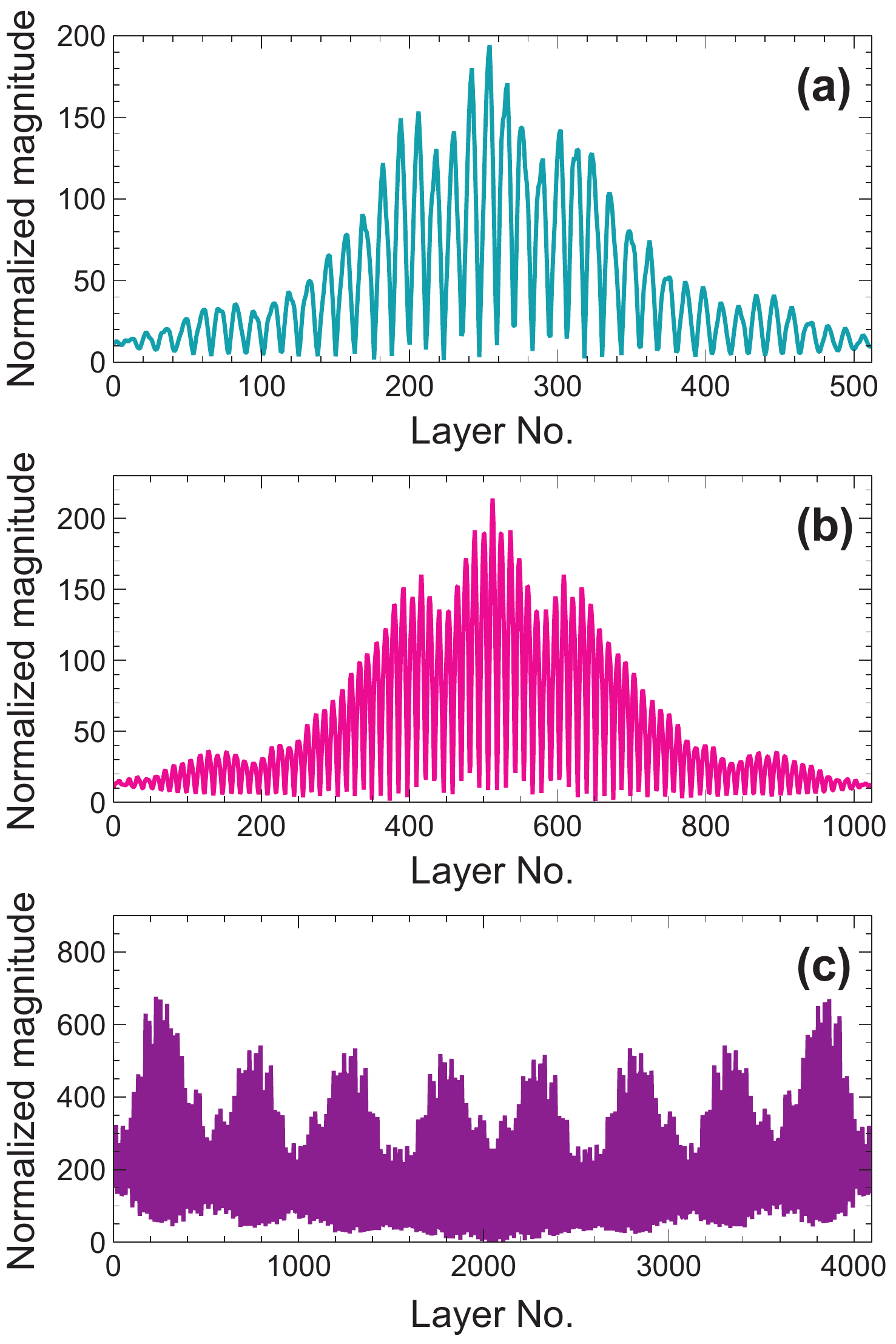}
	\caption{(a), (b), (c) Normalized (magnitude) field distributions of the quasi-localized states associated with representative transmittance peaks ($d/\lambda=0.079, 0.061, 0.085$, respectively) in the corresponding panels of Fig. \ref{Figure12}.}
	\label{Figure13}
\end{figure}

Next, we consider a scenario featuring small optical gain, namely, $\delta=-10^{-3}$.
Figure \ref{Figure12} shows some representative transmittance responses in the vicinity of fractal gaps, at different stages of growth, characterized by the presence of sharp peaks with very strong amplitudes (up to values of $\sim10^4$), which are indicative of lasing conditions. Also in these cases, as shown in Fig. \ref{Figure13}, the corresponding field distributions resemble quasi-localized states. Similar behaviors are also observed for the reflectance responses. Conversely, transmittance and reflectance for the EMT and periodic counterparts remain near-unit and very small, respectively. Qualitatively similar results (not shown for brevity) are also observed in connection with Fabry-P\'erot-type resonant modes excited nearby the critical incidence.

To better illustrate the difference between the observed response and the EMT prediction, we consider the lasing condition derived by enforcing a pole in the transmission (or reflection) coefficient for the EMT case (see Appendix \ref{Sec:AppB} for details)
\beq
\tan\left({\bar k}_z D\right)=\frac{2i{\bar k}_z k_{ze}}{{\bar k}_z^2+k_{ze}^2},
\label{eq:lasing}
\eeq
which can admit real-frequency solutions in the presence of gain. To give an idea, for the parameters corresponding to the resonant peak in Fig. \ref{Figure12}(b) ($n=10$, $\theta_i=31.6^o$, $d/\lambda=0.061$), the EMT prediction in Eq. (\ref{eq:lasing}) yields an overall thickness $D\sim2000\lambda$, i.e., more than 60 time thicker than the ThM case. Alternatively, for the same overall thickness $D=31.23\lambda$ as for the ThM case, the EMT prediction yields a gain coefficient $\delta=-0.066$, i.e., over 60 time larger.

The above results indicate that both types of aperiodic-order-induced nonlocality-enhancement mechanisms exhibit a remarkably high sensitivity to very small loss-gain values, which may find potential applications to optical sensing, absorbers and low-threshold lasers.

%-------------------------------------------------------------
\subsection{Some Remarks}
%-------------------------------------------------------------
A few remarks are in order on the assumptions and restrictions of our study. First, one may argue that the material parameters considered in the multilayers (especially $\varepsilon_a=1$) are not realistic for an experimental validation. As previously mentioned, the main motivation behind this parameter choice was to facilitate direct comparison with the results in previous studies on periodic and random scenarios \mbox{\cite{Sheinfux:2014sm,Castaldi:2018be,Sheinfux:2016cr,Lei:2017rt}}.
As also demonstrated by the experimental studies in Refs. \mbox{\cite{Zhukovsky:2015ed,Sheinfux:2017oo}}, the phenomena of interest remain visible when realistic materials (e.g., silica and titania) are instead utilized. However, the lower the material contrast, the more difficult the observation. For the periodic multilayer case, in Ref. \mbox{\cite{Castaldi:2018be}}, we were able to derive analytically the relationship between the material contrast and the critical size of the multilayer for which the breakdown phenomena could be observed. For the ThM case of interest here, an analytic study is not possible, but similar qualitative conclusions are expected to hold. Within this framework, another potentially critical aspect is the exponential increase of the multilayer size with the stage of growth $n$. In the examples shown, for the assumed parameters, the buildup effects leading to enhanced nonlocality turn out to occur for stages of growth $n\gtrsim 8$. While particularly high values of $n$ would clearly lead to technologically unfeasible structures, we remark that stages of growth around the threshold values $n=8$ and $n=9$ (i.e., hundreds of layers) are within reach for current nanofabrication technologies, as demonstrated in recent experimental studies \mbox{\cite{Sheinfux:2017oo}}.

Moreover, in connection with our assumption of TE polarization, once again to facilitate direct comparison with the results from previous related studies \mbox{\cite{Sheinfux:2014sm,Castaldi:2018be,Sheinfux:2016cr,Lei:2017rt}}, we note that enhanced nonlocality generally occurs for the transverse-magnetic polarization as well \mbox{\cite{Sheinfux:2014sm}}, although its visibility may be less pronounced \mbox{\cite{Zhukovsky:2015ed}}.

Finally, we remark that, unlike the periodic-multilayer case \mbox{\cite{Castaldi:2018be}}, it was not possible here to identify some closed-form parameters relating the optical response with the ThM aperiodic geometry, due to the more complex character of the arising trace and antitrace maps (not solvable analytically). Nevertheless, our parametric studies elucidate some representative mechanisms and effects that are distinctive of the ThM geometry.

%%%%%%%%%%%%%%%%%%%%%%%%%%%%%%%%%%%%%%%%%%%%%%%%%%%%%%%%%%%%%%
\section{Conclusions and Perspectives}
%%%%%%%%%%%%%%%%%%%%%%%%%%%%%%%%%%%%%%%%%%%%%%%%%%%%%%%%%%%%%%
\label{Sec:Conclusions}
To sum up, we have shown that, in aperiodically ordered, fully dielectric multilayered metamaterials based on the ThM geometry, the inherently weak nonlocality exhibited in the deeply subwavelength regime can be substantially enhanced via the buildup of boundary effects that are fundamentally different from those observed in the periodic case \cite{Sheinfux:2014sm,Zhukovsky:2015ed,Andryieuski:2015ae,Popov:2016oa,Lei:2017rt,Maurel:2018so,Castaldi:2018be}. These effects are manifested as strong departures of the optical response (reflectance and transmittance, as well as absorbance or lasing in the presence of small loss or gain, respectively)
from the EMT prediction and periodic counterpart, with distinctive footprints such as fractal gaps and quasi-localized states. 

We stress, once again, that the comparison with the periodic case is particularly meaningful, since the two geometries (ThM and periodic) contain {\em exactly the same} amounts of each of the material constituents, the only difference being the spatial order. This provides further evidence that, even at deeply subwavelength scales, spatial order may strongly affect the optical response. 

Our outcomes constitute a first step toward extending the previous studies on periodic \cite{Sheinfux:2014sm,Zhukovsky:2015ed,Andryieuski:2015ae,Popov:2016oa,Lei:2017rt,Maurel:2018so,Castaldi:2018be}
and randomly disordered \cite{Sheinfux:2016cr,Sheinfux:2017oo} 
geometries to the intermediate realm of ``orderly disorder'' and, albeit focused on a specific geometry, provide some generally applicable tools. As shown in Refs. \cite{Kolar:1990ma,Kolar:1990tm}, the trace and antitrace map formalism can in principle be applied to {\em generic} aperiodic sequences based on two-symbol substitution rules. Therefore, among the possible follow-up studies, it looks very intriguing to explore different aperiodically ordered geometries. 
For instance, it would be very interesting to explore to what extent some distinctive properties of the optical response
	of Fibonacci-type photonic quasycrystals (e.g., self-similarity in the spectrum, critical states of multifractal nature, etc.)  \mbox{\cite{Vardeny:2013hj}} are also observable in the deeply subwavelength regime.
Also of great interest are more application-oriented studies on the promising potentials that have emerged in connection with optical sensors, absorbers and lasers. Within this framework, we are currently pursuing a systematic study of the effects of enhanced nonlocality on the (bulk and surface) optical sensitivity response of ThM multilayers, which will be the subject of a forthcoming paper.

\appendix

%%%%%%%%%%%%%%%%%%%%%%%%%%%%%%%%%%%%%%%%%%%%%%%%%%%%%%%%%%%%%%
\section{Details on Eqs. (\ref{eq:tau}) and (\ref{eq:rhon})}
%%%%%%%%%%%%%%%%%%%%%%%%%%%%%%%%%%%%%%%%%%%%%%%%%%%%%%%%%%%%%%
\label{Sec:AppA}
In view of Eq. (\ref{eq:Ei}) and the definitions of the transmission and reflection coefficients in Eqs. (\ref{eq:taun}) and (\ref{eq:rhon}), respectively, the total electric fields at the input ($z=0$) and output ($z=D$) interfaces can be written as
\begin{subequations}
\begin{eqnarray}
E_y\left(x,z=0\right)&=&E_0\left(1+\rho_n\right)
\exp\left(ik_x x\right),\\
E_y\left(x,z=D\right)&=&E_0\tau_n
\exp\left(ik_x x\right).
\end{eqnarray}
\end{subequations}
By calculating the corresponding tangential magnetic fields from the relevant Maxwell's equation, the  matrix equation in Eq. (\ref{eq:Etot}) can be rewritten as
\beq
\left[
\begin{array}{cc}
	1+\rho_n\\
	-i\left(1-\rho_n\right)
\end{array}
\right]={\underline {\underline {\cal M}}}^{(n)}\cdot \left[
\begin{array}{cc}
	\tau_n\\
	-i\tau_n
\end{array}
\right],
\label{eq:TM1}
\eeq
from which Eqs. (\ref{eq:tau}) and (\ref{eq:rhon}) readily follow by solving the linear system of equations, and exploiting the unimodular character of the matrix.

%%%%%%%%%%%%%%%%%%%%%%%%%%%%%%%%%%%%%%%%%%%%%%%%%%%%%%%%%%%%%%
\section{Details on Eqs.  (\ref{eq:enh}) and (\ref{eq:lasing})}
%%%%%%%%%%%%%%%%%%%%%%%%%%%%%%%%%%%%%%%%%%%%%%%%%%%%%%%%%%%%%%
\label{Sec:AppB}
To calculate the EMT predictions, we consider a homogeneous slab of thickness $D$ and relative permittivity ${\bar \varepsilon}_{\parallel}$.
For given incidence conditions, the electric field inside the structure has the form of a standing wave
\beq
{\bar E}_y\left(z\right)={\bar E}^+
\exp\left[
i{\bar k}_z\left(z-D\right)
\right]
\left\{
1+{\bar \Gamma}
\exp\left[
-2i{\bar k}_z\left(z-D\right)
\right]
\right\},
\label{eq:EEy}
\eeq
where an irrelevant $\exp\left(ik_xx\right)$ term is omitted, ${\bar E}^+$ is a complex amplitude to be determined,
\beq
{\bar \Gamma}=\frac{Z_e-{\bar Z}}{Z_e+{\bar Z}}
\label{eq:Gamma}
\eeq
is the partial reflection coefficient between the exterior and effective media, and
\beq
{\bar Z}=\frac{\omega \mu_0}{{\bar k}_z}
\eeq
is the TE wave impedance of the effective medium.
 In the exterior region $z<0$, the total field is obtained by summing the incident [see Eq. (\ref{eq:Ei})] and reflected contributions, viz.,
\beq
	{\bar E}_y\left(z\right)=E_0
	\left[
	1+{\bar \rho}
	\exp\left(
	-2ik_{ze}z
	\right)
	\right],
\eeq	
where ${\bar \rho}$ can be obtained from Eq. (\ref{eq:rhon}) by assuming a single layer of thickness $D$ and relative permittivity ${\bar \varepsilon}_{\parallel}$.
By enforcing the continuity of the electric field at the interface $z=0$, after some algebra, we obtain
\beq
{\bar E}^+=\frac{2E_0{\bar Z}\left(Z_e+{\bar Z}\right)\exp\left(-2i{\bar k}_zD\right)}
{\exp\left(-2i{\bar k}_zD\right)\left(Z_e+{\bar Z}\right)^2-\left(Z_e-{\bar Z}\right)^2}.
\eeq
By recalling that, in view of the assumed parameters, $\varepsilon_e>{\bar \varepsilon}_{\parallel}$ and hence $Z_e<{\bar Z}$, we observe from
Eq. (\ref{eq:Gamma}) that ${\bar \Gamma}<0$. Therefore, by assuming the slab thicker than half a wavelength, it follows from Eq. (\ref{eq:EEy}) that
\beq
{\bar \gamma}=\frac{\left|{\bar E}^+\right|\left(1-{\bar \Gamma}\right)}{E_0}=\frac{4{\bar Z}^2}
{\left(Z_e+{\bar Z}\right)^2-\left(Z_e-{\bar Z}\right)^2}=\frac{{\bar Z}}{Z_e}=\frac{k_{ze}}{{\bar k}_z},
\eeq
which corresponds to the result in Eq. (\ref{eq:enh}).

Likewise, from Eq. (\ref{eq:tau}), the EMT prediction of the transmission coefficient is
\beq
{\bar \tau}=\frac{1}{\cos\left({\bar k}_z D\right)-i\left(\displaystyle{\frac{{\bar k}_z}{k_{ze}}+\frac{k_{ze}}{{\bar k}_z}}\right)
	\sin\left({\bar k}_z D\right)},
\eeq
from which the lasing condition in Eq. (\ref{eq:lasing}) directly follows by zeroing the denominator.

%\bibliographystyle{apsrev4-1}
%\bibliography{ThM-ML_breakdown}

\begin{thebibliography}{49}%
	\makeatletter
	\providecommand \@ifxundefined [1]{%
		\@ifx{#1\undefined}
	}%
	\providecommand \@ifnum [1]{%
		\ifnum #1\expandafter \@firstoftwo
		\else \expandafter \@secondoftwo
		\fi
	}%
	\providecommand \@ifx [1]{%
		\ifx #1\expandafter \@firstoftwo
		\else \expandafter \@secondoftwo
		\fi
	}%
	\providecommand \natexlab [1]{#1}%
	\providecommand \enquote  [1]{``#1''}%
	\providecommand \bibnamefont  [1]{#1}%
	\providecommand \bibfnamefont [1]{#1}%
	\providecommand \citenamefont [1]{#1}%
	\providecommand \href@noop [0]{\@secondoftwo}%
	\providecommand \href [0]{\begingroup \@sanitize@url \@href}%
	\providecommand \@href[1]{\@@startlink{#1}\@@href}%
	\providecommand \@@href[1]{\endgroup#1\@@endlink}%
	\providecommand \@sanitize@url [0]{\catcode `\\12\catcode `\$12\catcode
		`\&12\catcode `\#12\catcode `\^12\catcode `\_12\catcode `\%12\relax}%
	\providecommand \@@startlink[1]{}%
	\providecommand \@@endlink[0]{}%
	\providecommand \url  [0]{\begingroup\@sanitize@url \@url }%
	\providecommand \@url [1]{\endgroup\@href {#1}{\urlprefix }}%
	\providecommand \urlprefix  [0]{URL }%
	\providecommand \Eprint [0]{\href }%
	\providecommand \doibase [0]{http://dx.doi.org/}%
	\providecommand \selectlanguage [0]{\@gobble}%
	\providecommand \bibinfo  [0]{\@secondoftwo}%
	\providecommand \bibfield  [0]{\@secondoftwo}%
	\providecommand \translation [1]{[#1]}%
	\providecommand \BibitemOpen [0]{}%
	\providecommand \bibitemStop [0]{}%
	\providecommand \bibitemNoStop [0]{.\EOS\space}%
	\providecommand \EOS [0]{\spacefactor3000\relax}%
	\providecommand \BibitemShut  [1]{\csname bibitem#1\endcsname}%
	\let\auto@bib@innerbib\@empty
	%</preamble>
	\bibitem [{\citenamefont {Capolino}(2009)}]{Capolino:2009vr}%
	\BibitemOpen
	\bibfield  {author} {\bibinfo {author} {\bibfnamefont {F.}~\bibnamefont
			{Capolino}},\ }\href@noop {} {\emph {\bibinfo {title} {{Theory and Phenomena
					of Metamaterials}}}}\ (\bibinfo  {publisher} {CRC Press},\ \bibinfo {address}
	{Boca Raton, FL},\ \bibinfo {year} {2009})\BibitemShut {NoStop}%
	\bibitem [{\citenamefont {Cai}\ and\ \citenamefont
		{Shalaev}(2010)}]{Cai:2010om}%
	\BibitemOpen
	\bibfield  {author} {\bibinfo {author} {\bibfnamefont {W.}~\bibnamefont
			{Cai}}\ and\ \bibinfo {author} {\bibfnamefont {V.~M.}\ \bibnamefont
			{Shalaev}},\ }\href@noop {} {\emph {\bibinfo {title} {{Optical Metamaterials:
					Fundamentals and Applications}}}}\ (\bibinfo  {publisher} {Springer},\
	\bibinfo {address} {New York},\ \bibinfo {year} {2010})\BibitemShut {NoStop}%
	\bibitem [{\citenamefont {Urbas}\ \emph {et~al.}(2016)\citenamefont {Urbas},
		\citenamefont {Jacob}, \citenamefont {Dal~Negro}, \citenamefont {Engheta},
		\citenamefont {Boardman}, \citenamefont {Egan}, \citenamefont {Khanikaev},
		\citenamefont {Menon}, \citenamefont {Ferrera}, \citenamefont {Kinsey},
		\citenamefont {DeVault}, \citenamefont {Kim}, \citenamefont {Shalaev},
		\citenamefont {Boltasseva}, \citenamefont {Valentine}, \citenamefont
		{Pfeiffer}, \citenamefont {Grbic}, \citenamefont {Narimanov}, \citenamefont
		{Zhu}, \citenamefont {Fan}, \citenamefont {Al\`u}, \citenamefont {Poutrina},
		\citenamefont {Litchinitser}, \citenamefont {Noginov}, \citenamefont
		{MacDonald}, \citenamefont {Plum}, \citenamefont {Liu}, \citenamefont
		{Nealey}, \citenamefont {Kagan}, \citenamefont {Murray}, \citenamefont
		{Pawlak}, \citenamefont {Smolyaninov}, \citenamefont {Smolyaninova},\ and\
		\citenamefont {Chanda}}]{Urbas:2016ro}%
	\BibitemOpen
	\bibfield  {author} {\bibinfo {author} {\bibfnamefont {A.~M.}\ \bibnamefont
			{Urbas}}, \bibinfo {author} {\bibfnamefont {Z.}~\bibnamefont {Jacob}},
		\bibinfo {author} {\bibfnamefont {L.}~\bibnamefont {Dal~Negro}}, \bibinfo
		{author} {\bibfnamefont {N.}~\bibnamefont {Engheta}}, \bibinfo {author}
		{\bibfnamefont {A.~D.}\ \bibnamefont {Boardman}}, \bibinfo {author}
		{\bibfnamefont {P.}~\bibnamefont {Egan}}, \bibinfo {author} {\bibfnamefont
			{A.~B.}\ \bibnamefont {Khanikaev}}, \bibinfo {author} {\bibfnamefont
			{V.}~\bibnamefont {Menon}}, \bibinfo {author} {\bibfnamefont
			{M.}~\bibnamefont {Ferrera}}, \bibinfo {author} {\bibfnamefont
			{N.}~\bibnamefont {Kinsey}}, \bibinfo {author} {\bibfnamefont
			{C.}~\bibnamefont {DeVault}}, \bibinfo {author} {\bibfnamefont
			{J.}~\bibnamefont {Kim}}, \bibinfo {author} {\bibfnamefont {V.}~\bibnamefont
			{Shalaev}}, \bibinfo {author} {\bibfnamefont {A.}~\bibnamefont {Boltasseva}},
		\bibinfo {author} {\bibfnamefont {J.}~\bibnamefont {Valentine}}, \bibinfo
		{author} {\bibfnamefont {C.}~\bibnamefont {Pfeiffer}}, \bibinfo {author}
		{\bibfnamefont {A.}~\bibnamefont {Grbic}}, \bibinfo {author} {\bibfnamefont
			{E.}~\bibnamefont {Narimanov}}, \bibinfo {author} {\bibfnamefont
			{L.}~\bibnamefont {Zhu}}, \bibinfo {author} {\bibfnamefont {S.}~\bibnamefont
			{Fan}}, \bibinfo {author} {\bibfnamefont {A.}~\bibnamefont {Al\`u}}, \bibinfo
		{author} {\bibfnamefont {E.}~\bibnamefont {Poutrina}}, \bibinfo {author}
		{\bibfnamefont {N.~M.}\ \bibnamefont {Litchinitser}}, \bibinfo {author}
		{\bibfnamefont {M.~A.}\ \bibnamefont {Noginov}}, \bibinfo {author}
		{\bibfnamefont {K.~F.}\ \bibnamefont {MacDonald}}, \bibinfo {author}
		{\bibfnamefont {E.}~\bibnamefont {Plum}}, \bibinfo {author} {\bibfnamefont
			{X.}~\bibnamefont {Liu}}, \bibinfo {author} {\bibfnamefont {P.~F.}\
			\bibnamefont {Nealey}}, \bibinfo {author} {\bibfnamefont {C.~R.}\
			\bibnamefont {Kagan}}, \bibinfo {author} {\bibfnamefont {C.~B.}\ \bibnamefont
			{Murray}}, \bibinfo {author} {\bibfnamefont {D.~A.}\ \bibnamefont {Pawlak}},
		\bibinfo {author} {\bibfnamefont {I.~I.}\ \bibnamefont {Smolyaninov}},
		\bibinfo {author} {\bibfnamefont {V.~N.}\ \bibnamefont {Smolyaninova}}, \
		and\ \bibinfo {author} {\bibfnamefont {D.}~\bibnamefont {Chanda}},\
	}\href@noop {} {\bibfield  {journal} {\bibinfo  {journal} {J. Opt.}\ }\textbf
		{\bibinfo {volume} {18}},\ \bibinfo {pages} {093005} (\bibinfo {year}
		{2016})}\BibitemShut {NoStop}%
	\bibitem [{\citenamefont {Joannopoulos}\ \emph {et~al.}(2008)\citenamefont
		{Joannopoulos}, \citenamefont {Johnson}, \citenamefont {Winn},\ and\
		\citenamefont {Meade}}]{Joannopoulos:2008pc}%
	\BibitemOpen
	\bibfield  {author} {\bibinfo {author} {\bibfnamefont {J.~D.}\ \bibnamefont
			{Joannopoulos}}, \bibinfo {author} {\bibfnamefont {S.~G.}\ \bibnamefont
			{Johnson}}, \bibinfo {author} {\bibfnamefont {J.~N.}\ \bibnamefont {Winn}}, \
		and\ \bibinfo {author} {\bibfnamefont {R.~D.}\ \bibnamefont {Meade}},\
	}\href@noop {} {\emph {\bibinfo {title} {Photonic Crystals: Molding the Flow
				of Light}}},\ \bibinfo {edition} {2nd}\ ed.\ (\bibinfo  {publisher}
	{Princeton University Press},\ \bibinfo {address} {Princeton, NJ, USA},\
	\bibinfo {year} {2008})\BibitemShut {NoStop}%
	\bibitem [{\citenamefont {Smith}\ and\ \citenamefont
		{Pendry}(2006)}]{Smith:2006ho}%
	\BibitemOpen
	\bibfield  {author} {\bibinfo {author} {\bibfnamefont {D.~R.}\ \bibnamefont
			{Smith}}\ and\ \bibinfo {author} {\bibfnamefont {J.~B.}\ \bibnamefont
			{Pendry}},\ }\href {\doibase 10.1364/JOSAB.23.000391} {\bibfield  {journal}
		{\bibinfo  {journal} {J. Opt. Soc. Am. B}\ }\textbf {\bibinfo {volume}
			{23}},\ \bibinfo {pages} {391} (\bibinfo {year} {2006})}\BibitemShut
	{NoStop}%
	\bibitem [{\citenamefont {Smith}\ \emph {et~al.}(2005)\citenamefont {Smith},
		\citenamefont {Vier}, \citenamefont {Koschny},\ and\ \citenamefont
		{Soukoulis}}]{Smith:2005ep}%
	\BibitemOpen
	\bibfield  {author} {\bibinfo {author} {\bibfnamefont {D.~R.}\ \bibnamefont
			{Smith}}, \bibinfo {author} {\bibfnamefont {D.~C.}\ \bibnamefont {Vier}},
		\bibinfo {author} {\bibfnamefont {T.}~\bibnamefont {Koschny}}, \ and\
		\bibinfo {author} {\bibfnamefont {C.~M.}\ \bibnamefont {Soukoulis}},\ }\href
	{\doibase 10.1103/PhysRevE.71.036617} {\bibfield  {journal} {\bibinfo
			{journal} {Phys. Rev. E}\ }\textbf {\bibinfo {volume} {71}},\ \bibinfo
		{pages} {036617} (\bibinfo {year} {2005})}\BibitemShut {NoStop}%
	\bibitem [{\citenamefont {Arslanagi\'c}\ \emph {et~al.}(2013)\citenamefont
		{Arslanagi\'c}, \citenamefont {Hansen}, \citenamefont {Mortensen},
		\citenamefont {Gregersen}, \citenamefont {Sigmund}, \citenamefont
		{Ziolkowski},\ and\ \citenamefont {Breinbjerg}}]{Arslanagic:2013ar}%
	\BibitemOpen
	\bibfield  {author} {\bibinfo {author} {\bibfnamefont {S.}~\bibnamefont
			{Arslanagi\'c}}, \bibinfo {author} {\bibfnamefont {T.~V.}\ \bibnamefont
			{Hansen}}, \bibinfo {author} {\bibfnamefont {N.~A.}\ \bibnamefont
			{Mortensen}}, \bibinfo {author} {\bibfnamefont {A.~H.}\ \bibnamefont
			{Gregersen}}, \bibinfo {author} {\bibfnamefont {O.}~\bibnamefont {Sigmund}},
		\bibinfo {author} {\bibfnamefont {R.~W.}\ \bibnamefont {Ziolkowski}}, \ and\
		\bibinfo {author} {\bibfnamefont {O.}~\bibnamefont {Breinbjerg}},\ }\href
	{\doibase 10.1109/MAP.2013.6529320} {\bibfield  {journal} {\bibinfo
			{journal} {IEEE Antennas Propagation Mag.}\ }\textbf {\bibinfo {volume}
			{55}},\ \bibinfo {pages} {91} (\bibinfo {year} {2013})}\BibitemShut {NoStop}%
	\bibitem [{\citenamefont {Sihvola}(1999)}]{Sihvola:1999em}%
	\BibitemOpen
	\bibfield  {author} {\bibinfo {author} {\bibfnamefont {A.}~\bibnamefont
			{Sihvola}},\ }\href@noop {} {\emph {\bibinfo {title} {Electromagnetic Mixing
				Formulas and Applications}}},\ Electromagnetics and Radar Series\ (\bibinfo
	{publisher} {IET},\ \bibinfo {address} {London, UK},\ \bibinfo {year}
	{1999})\BibitemShut {NoStop}%
	\bibitem [{\citenamefont {Landau}\ and\ \citenamefont
		{Lifshitz}(1960)}]{Landau:1960eo}%
	\BibitemOpen
	\bibfield  {author} {\bibinfo {author} {\bibfnamefont {D.}~\bibnamefont
			{Landau}}\ and\ \bibinfo {author} {\bibfnamefont {E.~M.}\ \bibnamefont
			{Lifshitz}},\ }\href@noop {} {\emph {\bibinfo {title} {Electrodynamics of
				Continuous Media}}}\ (\bibinfo  {publisher} {Pergamon Press},\ \bibinfo
	{address} {New York},\ \bibinfo {year} {1960})\BibitemShut {NoStop}%
	\bibitem [{\citenamefont {Agranovich}\ and\ \citenamefont
		{Ginzburg}(2013)}]{Agranovich:2013co}%
	\BibitemOpen
	\bibfield  {author} {\bibinfo {author} {\bibfnamefont {V.}~\bibnamefont
			{Agranovich}}\ and\ \bibinfo {author} {\bibfnamefont {V.}~\bibnamefont
			{Ginzburg}},\ }\href@noop {} {\emph {\bibinfo {title} {Crystal Optics with
				Spatial Dispersion, and Excitons}}},\ Springer Series in Solid-State
	Sciences\ (\bibinfo  {publisher} {Springer-Vergal},\ \bibinfo {address}
	{Berlin/Heidelberg},\ \bibinfo {year} {2013})\BibitemShut {NoStop}%
	\bibitem [{\citenamefont {Silveirinha}(2007)}]{Silveirinha:2007mh}%
	\BibitemOpen
	\bibfield  {author} {\bibinfo {author} {\bibfnamefont {M.~G.}\ \bibnamefont
			{Silveirinha}},\ }\href {\doibase 10.1103/PhysRevB.75.115104} {\bibfield
		{journal} {\bibinfo  {journal} {Phys. Rev. B}\ }\textbf {\bibinfo {volume}
			{75}},\ \bibinfo {pages} {115104} (\bibinfo {year} {2007})}\BibitemShut
	{NoStop}%
	\bibitem [{\citenamefont {Al\`u}(2011)}]{Alu:2011fp}%
	\BibitemOpen
	\bibfield  {author} {\bibinfo {author} {\bibfnamefont {A.}~\bibnamefont
			{Al\`u}},\ }\href {\doibase 10.1103/PhysRevB.84.075153} {\bibfield  {journal}
		{\bibinfo  {journal} {Phys. Rev. B}\ }\textbf {\bibinfo {volume} {84}},\
		\bibinfo {pages} {075153} (\bibinfo {year} {2011})}\BibitemShut {NoStop}%
	\bibitem [{\citenamefont {Ciattoni}\ and\ \citenamefont
		{Rizza}(2015)}]{Ciattoni:2015nh}%
	\BibitemOpen
	\bibfield  {author} {\bibinfo {author} {\bibfnamefont {A.}~\bibnamefont
			{Ciattoni}}\ and\ \bibinfo {author} {\bibfnamefont {C.}~\bibnamefont
			{Rizza}},\ }\href {\doibase 10.1103/PhysRevB.91.184207} {\bibfield  {journal}
		{\bibinfo  {journal} {Phys. Rev. B}\ }\textbf {\bibinfo {volume} {91}},\
		\bibinfo {pages} {184207} (\bibinfo {year} {2015})}\BibitemShut {NoStop}%
	\bibitem [{\citenamefont {Elser}\ \emph {et~al.}(2007)\citenamefont {Elser},
		\citenamefont {Podolskiy}, \citenamefont {Salakhutdinov},\ and\ \citenamefont
		{Avrutsky}}]{Elser:2007ne}%
	\BibitemOpen
	\bibfield  {author} {\bibinfo {author} {\bibfnamefont {J.}~\bibnamefont
			{Elser}}, \bibinfo {author} {\bibfnamefont {V.~A.}\ \bibnamefont
			{Podolskiy}}, \bibinfo {author} {\bibfnamefont {I.}~\bibnamefont
			{Salakhutdinov}}, \ and\ \bibinfo {author} {\bibfnamefont {I.}~\bibnamefont
			{Avrutsky}},\ }\href {\doibase 10.1063/1.2737935} {\bibfield  {journal}
		{\bibinfo  {journal} {Appl. Phys. Lett.}\ }\textbf {\bibinfo {volume} {90}},\
		\bibinfo {pages} {191109} (\bibinfo {year} {2007})}\BibitemShut {NoStop}%
	\bibitem [{\citenamefont {Chebykin}\ \emph {et~al.}(2011)\citenamefont
		{Chebykin}, \citenamefont {Orlov}, \citenamefont {Vozianova}, \citenamefont
		{Maslovski}, \citenamefont {Kivshar},\ and\ \citenamefont
		{Belov}}]{Chebykin:2011ne}%
	\BibitemOpen
	\bibfield  {author} {\bibinfo {author} {\bibfnamefont {A.~V.}\ \bibnamefont
			{Chebykin}}, \bibinfo {author} {\bibfnamefont {A.~A.}\ \bibnamefont {Orlov}},
		\bibinfo {author} {\bibfnamefont {A.~V.}\ \bibnamefont {Vozianova}}, \bibinfo
		{author} {\bibfnamefont {S.~I.}\ \bibnamefont {Maslovski}}, \bibinfo {author}
		{\bibfnamefont {Y.~S.}\ \bibnamefont {Kivshar}}, \ and\ \bibinfo {author}
		{\bibfnamefont {P.~A.}\ \bibnamefont {Belov}},\ }\href {\doibase
		10.1103/PhysRevB.84.115438} {\bibfield  {journal} {\bibinfo  {journal} {Phys.
				Rev. B}\ }\textbf {\bibinfo {volume} {84}},\ \bibinfo {pages} {115438}
		(\bibinfo {year} {2011})}\BibitemShut {NoStop}%
	\bibitem [{\citenamefont {Chebykin}\ \emph {et~al.}(2012)\citenamefont
		{Chebykin}, \citenamefont {Orlov}, \citenamefont {Simovski}, \citenamefont
		{Kivshar},\ and\ \citenamefont {Belov}}]{Chebykin:2012ne}%
	\BibitemOpen
	\bibfield  {author} {\bibinfo {author} {\bibfnamefont {A.~V.}\ \bibnamefont
			{Chebykin}}, \bibinfo {author} {\bibfnamefont {A.~A.}\ \bibnamefont {Orlov}},
		\bibinfo {author} {\bibfnamefont {C.~R.}\ \bibnamefont {Simovski}}, \bibinfo
		{author} {\bibfnamefont {Y.~S.}\ \bibnamefont {Kivshar}}, \ and\ \bibinfo
		{author} {\bibfnamefont {P.~A.}\ \bibnamefont {Belov}},\ }\href {\doibase
		10.1103/PhysRevB.86.115420} {\bibfield  {journal} {\bibinfo  {journal} {Phys.
				Rev. B}\ }\textbf {\bibinfo {volume} {86}},\ \bibinfo {pages} {115420}
		(\bibinfo {year} {2012})}\BibitemShut {NoStop}%
	\bibitem [{\citenamefont {Chern}(2013)}]{Chern:2013sd}%
	\BibitemOpen
	\bibfield  {author} {\bibinfo {author} {\bibfnamefont {R.-L.}\ \bibnamefont
			{Chern}},\ }\href {\doibase 10.1364/OE.21.016514} {\bibfield  {journal}
		{\bibinfo  {journal} {Opt. Express}\ }\textbf {\bibinfo {volume} {21}},\
		\bibinfo {pages} {16514} (\bibinfo {year} {2013})}\BibitemShut {NoStop}%
	\bibitem [{\citenamefont {Maier}(2007)}]{Maier:2007pf}%
	\BibitemOpen
	\bibfield  {author} {\bibinfo {author} {\bibfnamefont {S.}~\bibnamefont
			{Maier}},\ }\href@noop {} {\emph {\bibinfo {title} {Plasmonics: Fundamentals
				and Applications}}}\ (\bibinfo  {publisher} {Springer-Verlag},\ \bibinfo
	{address} {New York, NY, USA},\ \bibinfo {year} {2007})\BibitemShut {NoStop}%
	\bibitem [{\citenamefont {Orlov}\ \emph {et~al.}(2011)\citenamefont {Orlov},
		\citenamefont {Voroshilov}, \citenamefont {Belov},\ and\ \citenamefont
		{Kivshar}}]{Orlov:2011eo}%
	\BibitemOpen
	\bibfield  {author} {\bibinfo {author} {\bibfnamefont {A.~A.}\ \bibnamefont
			{Orlov}}, \bibinfo {author} {\bibfnamefont {P.~M.}\ \bibnamefont
			{Voroshilov}}, \bibinfo {author} {\bibfnamefont {P.~A.}\ \bibnamefont
			{Belov}}, \ and\ \bibinfo {author} {\bibfnamefont {Y.~S.}\ \bibnamefont
			{Kivshar}},\ }\href {\doibase 10.1103/PhysRevB.84.045424} {\bibfield
		{journal} {\bibinfo  {journal} {Phys. Rev. B}\ }\textbf {\bibinfo {volume}
			{84}},\ \bibinfo {pages} {045424} (\bibinfo {year} {2011})}\BibitemShut
	{NoStop}%
	\bibitem [{\citenamefont {Herzig~Sheinfux}\ \emph {et~al.}(2014)\citenamefont
		{Herzig~Sheinfux}, \citenamefont {Kaminer}, \citenamefont {Plotnik},
		\citenamefont {Bartal},\ and\ \citenamefont {Segev}}]{Sheinfux:2014sm}%
	\BibitemOpen
	\bibfield  {author} {\bibinfo {author} {\bibfnamefont {H.}~\bibnamefont
			{Herzig~Sheinfux}}, \bibinfo {author} {\bibfnamefont {I.}~\bibnamefont
			{Kaminer}}, \bibinfo {author} {\bibfnamefont {Y.}~\bibnamefont {Plotnik}},
		\bibinfo {author} {\bibfnamefont {G.}~\bibnamefont {Bartal}}, \ and\ \bibinfo
		{author} {\bibfnamefont {M.}~\bibnamefont {Segev}},\ }\href {\doibase
		10.1103/PhysRevLett.113.243901} {\bibfield  {journal} {\bibinfo  {journal}
			{Phys. Rev. Lett.}\ }\textbf {\bibinfo {volume} {113}},\ \bibinfo {pages}
		{243901} (\bibinfo {year} {2014})}\BibitemShut {NoStop}%
	\bibitem [{\citenamefont {Zhukovsky}\ \emph {et~al.}(2015)\citenamefont
		{Zhukovsky}, \citenamefont {Andryieuski}, \citenamefont {Takayama},
		\citenamefont {Shkondin}, \citenamefont {Malureanu}, \citenamefont {Jensen},\
		and\ \citenamefont {Lavrinenko}}]{Zhukovsky:2015ed}%
	\BibitemOpen
	\bibfield  {author} {\bibinfo {author} {\bibfnamefont {S.~V.}\ \bibnamefont
			{Zhukovsky}}, \bibinfo {author} {\bibfnamefont {A.}~\bibnamefont
			{Andryieuski}}, \bibinfo {author} {\bibfnamefont {O.}~\bibnamefont
			{Takayama}}, \bibinfo {author} {\bibfnamefont {E.}~\bibnamefont {Shkondin}},
		\bibinfo {author} {\bibfnamefont {R.}~\bibnamefont {Malureanu}}, \bibinfo
		{author} {\bibfnamefont {F.}~\bibnamefont {Jensen}}, \ and\ \bibinfo {author}
		{\bibfnamefont {A.~V.}\ \bibnamefont {Lavrinenko}},\ }\href {\doibase
		10.1103/PhysRevLett.115.177402} {\bibfield  {journal} {\bibinfo  {journal}
			{Phys. Rev. Lett.}\ }\textbf {\bibinfo {volume} {115}},\ \bibinfo {pages}
		{177402} (\bibinfo {year} {2015})}\BibitemShut {NoStop}%
	\bibitem [{\citenamefont {Andryieuski}\ \emph {et~al.}(2015)\citenamefont
		{Andryieuski}, \citenamefont {Lavrinenko},\ and\ \citenamefont
		{Zhukovsky}}]{Andryieuski:2015ae}%
	\BibitemOpen
	\bibfield  {author} {\bibinfo {author} {\bibfnamefont {A.}~\bibnamefont
			{Andryieuski}}, \bibinfo {author} {\bibfnamefont {A.~V.}\ \bibnamefont
			{Lavrinenko}}, \ and\ \bibinfo {author} {\bibfnamefont {S.~V.}\ \bibnamefont
			{Zhukovsky}},\ }\href@noop {} {\bibfield  {journal} {\bibinfo  {journal}
			{Nanotechnology}\ }\textbf {\bibinfo {volume} {26}},\ \bibinfo {pages}
		{184001} (\bibinfo {year} {2015})}\BibitemShut {NoStop}%
	\bibitem [{\citenamefont {Popov}\ \emph {et~al.}(2016)\citenamefont {Popov},
		\citenamefont {Lavrinenko},\ and\ \citenamefont {Novitsky}}]{Popov:2016oa}%
	\BibitemOpen
	\bibfield  {author} {\bibinfo {author} {\bibfnamefont {V.}~\bibnamefont
			{Popov}}, \bibinfo {author} {\bibfnamefont {A.~V.}\ \bibnamefont
			{Lavrinenko}}, \ and\ \bibinfo {author} {\bibfnamefont {A.}~\bibnamefont
			{Novitsky}},\ }\href {\doibase 10.1103/PhysRevB.94.085428} {\bibfield
		{journal} {\bibinfo  {journal} {Phys. Rev. B}\ }\textbf {\bibinfo {volume}
			{94}},\ \bibinfo {pages} {085428} (\bibinfo {year} {2016})}\BibitemShut
	{NoStop}%
	\bibitem [{\citenamefont {Lei}\ \emph {et~al.}(2017)\citenamefont {Lei},
		\citenamefont {Mao}, \citenamefont {Lu},\ and\ \citenamefont
		{Wang}}]{Lei:2017rt}%
	\BibitemOpen
	\bibfield  {author} {\bibinfo {author} {\bibfnamefont {X.}~\bibnamefont
			{Lei}}, \bibinfo {author} {\bibfnamefont {L.}~\bibnamefont {Mao}}, \bibinfo
		{author} {\bibfnamefont {Y.}~\bibnamefont {Lu}}, \ and\ \bibinfo {author}
		{\bibfnamefont {P.}~\bibnamefont {Wang}},\ }\href {\doibase
		10.1103/PhysRevB.96.035439} {\bibfield  {journal} {\bibinfo  {journal} {Phys.
				Rev. B}\ }\textbf {\bibinfo {volume} {96}},\ \bibinfo {pages} {035439}
		(\bibinfo {year} {2017})}\BibitemShut {NoStop}%
	\bibitem [{\citenamefont {Maurel}\ and\ \citenamefont
		{Marigo}(2018)}]{Maurel:2018so}%
	\BibitemOpen
	\bibfield  {author} {\bibinfo {author} {\bibfnamefont {A.}~\bibnamefont
			{Maurel}}\ and\ \bibinfo {author} {\bibfnamefont {J.-J.}\ \bibnamefont
			{Marigo}},\ }\href {\doibase 10.1103/PhysRevB.98.024306} {\bibfield
		{journal} {\bibinfo  {journal} {Phys. Rev. B}\ }\textbf {\bibinfo {volume}
			{98}},\ \bibinfo {pages} {024306} (\bibinfo {year} {2018})}\BibitemShut
	{NoStop}%
	\bibitem [{\citenamefont {Castaldi}\ \emph {et~al.}(2018)\citenamefont
		{Castaldi}, \citenamefont {Al\`u},\ and\ \citenamefont
		{Galdi}}]{Castaldi:2018be}%
	\BibitemOpen
	\bibfield  {author} {\bibinfo {author} {\bibfnamefont {G.}~\bibnamefont
			{Castaldi}}, \bibinfo {author} {\bibfnamefont {A.}~\bibnamefont {Al\`u}}, \
		and\ \bibinfo {author} {\bibfnamefont {V.}~\bibnamefont {Galdi}},\ }\href
	{\doibase 10.1103/PhysRevApplied.10.034060} {\bibfield  {journal} {\bibinfo
			{journal} {Phys. Rev. Appl.}\ }\textbf {\bibinfo {volume} {10}},\ \bibinfo
		{pages} {034060} (\bibinfo {year} {2018})}\BibitemShut {NoStop}%
	\bibitem [{\citenamefont {Herzig~Sheinfux}\ \emph {et~al.}(2016)\citenamefont
		{Herzig~Sheinfux}, \citenamefont {Kaminer}, \citenamefont {Genack},\ and\
		\citenamefont {Segev}}]{Sheinfux:2016cr}%
	\BibitemOpen
	\bibfield  {author} {\bibinfo {author} {\bibfnamefont {H.}~\bibnamefont
			{Herzig~Sheinfux}}, \bibinfo {author} {\bibfnamefont {I.}~\bibnamefont
			{Kaminer}}, \bibinfo {author} {\bibfnamefont {A.~Z.}\ \bibnamefont {Genack}},
		\ and\ \bibinfo {author} {\bibfnamefont {M.}~\bibnamefont {Segev}},\
	}\href@noop {} {\bibfield  {journal} {\bibinfo  {journal} {Nat. Commun.}\
		}\textbf {\bibinfo {volume} {7}},\ \bibinfo {pages} {12927} (\bibinfo {year}
		{2016})}\BibitemShut {NoStop}%
	\bibitem [{\citenamefont {Herzig~Sheinfux}\ \emph {et~al.}(2017)\citenamefont
		{Herzig~Sheinfux}, \citenamefont {Lumer}, \citenamefont {Ankonina},
		\citenamefont {Genack}, \citenamefont {Bartal},\ and\ \citenamefont
		{Segev}}]{Sheinfux:2017oo}%
	\BibitemOpen
	\bibfield  {author} {\bibinfo {author} {\bibfnamefont {H.}~\bibnamefont
			{Herzig~Sheinfux}}, \bibinfo {author} {\bibfnamefont {Y.}~\bibnamefont
			{Lumer}}, \bibinfo {author} {\bibfnamefont {G.}~\bibnamefont {Ankonina}},
		\bibinfo {author} {\bibfnamefont {A.~Z.}\ \bibnamefont {Genack}}, \bibinfo
		{author} {\bibfnamefont {G.}~\bibnamefont {Bartal}}, \ and\ \bibinfo {author}
		{\bibfnamefont {M.}~\bibnamefont {Segev}},\ }\href {\doibase
		10.1126/science.aah6822} {\bibfield  {journal} {\bibinfo  {journal}
			{Science}\ }\textbf {\bibinfo {volume} {356}},\ \bibinfo {pages} {953}
		(\bibinfo {year} {2017})}\BibitemShut {NoStop}%
	\bibitem [{\citenamefont {Shechtman}\ \emph {et~al.}(1984)\citenamefont
		{Shechtman}, \citenamefont {Blech}, \citenamefont {Gratias},\ and\
		\citenamefont {Cahn}}]{Shechtman:1984mp}%
	\BibitemOpen
	\bibfield  {author} {\bibinfo {author} {\bibfnamefont {D.}~\bibnamefont
			{Shechtman}}, \bibinfo {author} {\bibfnamefont {I.}~\bibnamefont {Blech}},
		\bibinfo {author} {\bibfnamefont {D.}~\bibnamefont {Gratias}}, \ and\
		\bibinfo {author} {\bibfnamefont {J.~W.}\ \bibnamefont {Cahn}},\ }\href
	{\doibase 10.1103/PhysRevLett.53.1951} {\bibfield  {journal} {\bibinfo
			{journal} {Phys. Rev. Lett.}\ }\textbf {\bibinfo {volume} {53}},\ \bibinfo
		{pages} {1951} (\bibinfo {year} {1984})}\BibitemShut {NoStop}%
	\bibitem [{\citenamefont {Levine}\ and\ \citenamefont
		{Steinhardt}(1984)}]{Levine:1984qa}%
	\BibitemOpen
	\bibfield  {author} {\bibinfo {author} {\bibfnamefont {D.}~\bibnamefont
			{Levine}}\ and\ \bibinfo {author} {\bibfnamefont {P.~J.}\ \bibnamefont
			{Steinhardt}},\ }\href {\doibase 10.1103/PhysRevLett.53.2477} {\bibfield
		{journal} {\bibinfo  {journal} {Phys. Rev. Lett.}\ }\textbf {\bibinfo
			{volume} {53}},\ \bibinfo {pages} {2477} (\bibinfo {year}
		{1984})}\BibitemShut {NoStop}%
	\bibitem [{\citenamefont {Maci\'{a}}(2006)}]{Macia:2006tr}%
	\BibitemOpen
	\bibfield  {author} {\bibinfo {author} {\bibfnamefont {E.}~\bibnamefont
			{Maci\'{a}}},\ }\href@noop {} {\bibfield  {journal} {\bibinfo  {journal}
			{Rep. Progr. Phys.}\ }\textbf {\bibinfo {volume} {69}},\ \bibinfo {pages}
		{397} (\bibinfo {year} {2006})}\BibitemShut {NoStop}%
	\bibitem [{\citenamefont {Dal~Negro}\ and\ \citenamefont
		{Boriskina}(2011)}]{DalNegro:2011da}%
	\BibitemOpen
	\bibfield  {author} {\bibinfo {author} {\bibfnamefont {L.}~\bibnamefont
			{Dal~Negro}}\ and\ \bibinfo {author} {\bibfnamefont {S.}~\bibnamefont
			{Boriskina}},\ }\href {\doibase 10.1002/lpor.201000046} {\bibfield  {journal}
		{\bibinfo  {journal} {Laser Photonics Rev.}\ }\textbf {\bibinfo {volume}
			{6}},\ \bibinfo {pages} {178} (\bibinfo {year} {2011})}\BibitemShut {NoStop}%
	\bibitem [{\citenamefont {Queff{\'e}lec}(2010)}]{Queffelec:2010sd}%
	\BibitemOpen
	\bibfield  {author} {\bibinfo {author} {\bibfnamefont {M.}~\bibnamefont
			{Queff{\'e}lec}},\ }\href@noop {} {\emph {\bibinfo {title} {Substitution
				Dynamical Systems - Spectral Analysis}}},\ Lecture Notes in Mathematics\
	(\bibinfo  {publisher} {Springer-Verlag},\ \bibinfo {address}
	{Berlin/Heidelberg},\ \bibinfo {year} {2010})\BibitemShut {NoStop}%
	\bibitem [{\citenamefont {Liu}(1997)}]{Liu:1997po}%
	\BibitemOpen
	\bibfield  {author} {\bibinfo {author} {\bibfnamefont {N.-H.}\ \bibnamefont
			{Liu}},\ }\href {\doibase 10.1103/PhysRevB.55.3543} {\bibfield  {journal}
		{\bibinfo  {journal} {Phys. Rev. B}\ }\textbf {\bibinfo {volume} {55}},\
		\bibinfo {pages} {3543} (\bibinfo {year} {1997})}\BibitemShut {NoStop}%
	\bibitem [{\citenamefont {Qiu}\ \emph {et~al.}(2003)\citenamefont {Qiu},
		\citenamefont {Peng}, \citenamefont {Huang}, \citenamefont {Liu},
		\citenamefont {Wang}, \citenamefont {Hu},\ and\ \citenamefont
		{Jiang}}]{Qiu:2003rt}%
	\BibitemOpen
	\bibfield  {author} {\bibinfo {author} {\bibfnamefont {F.}~\bibnamefont
			{Qiu}}, \bibinfo {author} {\bibfnamefont {R.~W.}\ \bibnamefont {Peng}},
		\bibinfo {author} {\bibfnamefont {X.~Q.}\ \bibnamefont {Huang}}, \bibinfo
		{author} {\bibfnamefont {Y.~M.}\ \bibnamefont {Liu}}, \bibinfo {author}
		{\bibfnamefont {M.}~\bibnamefont {Wang}}, \bibinfo {author} {\bibfnamefont
			{A.}~\bibnamefont {Hu}}, \ and\ \bibinfo {author} {\bibfnamefont {S.~S.}\
			\bibnamefont {Jiang}},\ }\href@noop {} {\bibfield  {journal} {\bibinfo
			{journal} {Europhys. Lett.}\ }\textbf {\bibinfo {volume} {63}},\ \bibinfo
		{pages} {853} (\bibinfo {year} {2003})}\BibitemShut {NoStop}%
	\bibitem [{\citenamefont {Dal~Negro}\ \emph {et~al.}(2004)\citenamefont
		{Dal~Negro}, \citenamefont {Stolfi}, \citenamefont {Yi}, \citenamefont
		{Michel}, \citenamefont {Duan}, \citenamefont {Kimerling}, \citenamefont
		{LeBlanc},\ and\ \citenamefont {Haavisto}}]{DalNegro:2004pb}%
	\BibitemOpen
	\bibfield  {author} {\bibinfo {author} {\bibfnamefont {L.}~\bibnamefont
			{Dal~Negro}}, \bibinfo {author} {\bibfnamefont {M.}~\bibnamefont {Stolfi}},
		\bibinfo {author} {\bibfnamefont {Y.}~\bibnamefont {Yi}}, \bibinfo {author}
		{\bibfnamefont {J.}~\bibnamefont {Michel}}, \bibinfo {author} {\bibfnamefont
			{X.}~\bibnamefont {Duan}}, \bibinfo {author} {\bibfnamefont {L.~C.}\
			\bibnamefont {Kimerling}}, \bibinfo {author} {\bibfnamefont {J.}~\bibnamefont
			{LeBlanc}}, \ and\ \bibinfo {author} {\bibfnamefont {J.}~\bibnamefont
			{Haavisto}},\ }\href {\doibase 10.1063/1.1764602} {\bibfield  {journal}
		{\bibinfo  {journal} {Appl. Phys. Lett.}\ }\textbf {\bibinfo {volume} {84}},\
		\bibinfo {pages} {5186} (\bibinfo {year} {2004})}\BibitemShut {NoStop}%
	\bibitem [{\citenamefont {Jiang}\ \emph {et~al.}(2005)\citenamefont {Jiang},
		\citenamefont {Zhang}, \citenamefont {Feng}, \citenamefont {Huang},
		\citenamefont {Yi},\ and\ \citenamefont {Joannopoulos}}]{Jiang:2005pb}%
	\BibitemOpen
	\bibfield  {author} {\bibinfo {author} {\bibfnamefont {X.}~\bibnamefont
			{Jiang}}, \bibinfo {author} {\bibfnamefont {Y.}~\bibnamefont {Zhang}},
		\bibinfo {author} {\bibfnamefont {S.}~\bibnamefont {Feng}}, \bibinfo {author}
		{\bibfnamefont {K.~C.}\ \bibnamefont {Huang}}, \bibinfo {author}
		{\bibfnamefont {Y.}~\bibnamefont {Yi}}, \ and\ \bibinfo {author}
		{\bibfnamefont {J.~D.}\ \bibnamefont {Joannopoulos}},\ }\href {\doibase
		10.1063/1.1928317} {\bibfield  {journal} {\bibinfo  {journal} {Appl. Phys.
				Lett.}\ }\textbf {\bibinfo {volume} {86}},\ \bibinfo {pages} {201110}
		(\bibinfo {year} {2005})}\BibitemShut {NoStop}%
	\bibitem [{\citenamefont {Marianne~Hiltunen}\ and\ \citenamefont
		{Dal~Negro}(2008)}]{Hiltunen:2008lt}%
	\BibitemOpen
	\bibfield  {author} {\bibinfo {author} {\bibfnamefont {J.~M.}\ \bibnamefont
			{Marianne~Hiltunen}}\ and\ \bibinfo {author} {\bibfnamefont {L.}~\bibnamefont
			{Dal~Negro}},\ }in\ \href {\doibase 10.1117/12.781073} {\emph {\bibinfo
			{booktitle} {Proc. SPIE}}},\ Vol.\ \bibinfo {volume} {6989}\ (\bibinfo {year}
	{2008})\ pp.\ \bibinfo {pages} {6989--7}\BibitemShut {NoStop}%
	\bibitem [{\citenamefont {Grigoriev}\ and\ \citenamefont
		{Biancalana}(2010{\natexlab{a}})}]{Grigoriev:2010bm}%
	\BibitemOpen
	\bibfield  {author} {\bibinfo {author} {\bibfnamefont {V.}~\bibnamefont
			{Grigoriev}}\ and\ \bibinfo {author} {\bibfnamefont {F.}~\bibnamefont
			{Biancalana}},\ }\href@noop {} {\bibfield  {journal} {\bibinfo  {journal}
			{New J. Phys.}\ }\textbf {\bibinfo {volume} {12}},\ \bibinfo {pages} {053041}
		(\bibinfo {year} {2010}{\natexlab{a}})}\BibitemShut {NoStop}%
	\bibitem [{\citenamefont {Grigoriev}\ and\ \citenamefont
		{Biancalana}(2010{\natexlab{b}})}]{Grigoriev:2010bs}%
	\BibitemOpen
	\bibfield  {author} {\bibinfo {author} {\bibfnamefont {V.~V.}\ \bibnamefont
			{Grigoriev}}\ and\ \bibinfo {author} {\bibfnamefont {F.}~\bibnamefont
			{Biancalana}},\ }\href@noop {} {\bibfield  {journal} {\bibinfo  {journal}
			{Photonics Nanostruct. Fundam. Appl.}\ }\textbf {\bibinfo {volume} {8}},\
		\bibinfo {pages} {285} (\bibinfo {year} {2010}{\natexlab{b}})}\BibitemShut
	{NoStop}%
	\bibitem [{\citenamefont {Hsueh}\ \emph {et~al.}(2011)\citenamefont {Hsueh},
		\citenamefont {Wun}, \citenamefont {Lin},\ and\ \citenamefont
		{Cheng}}]{Hsueh:2011fo}%
	\BibitemOpen
	\bibfield  {author} {\bibinfo {author} {\bibfnamefont {W.~J.}\ \bibnamefont
			{Hsueh}}, \bibinfo {author} {\bibfnamefont {S.~J.}\ \bibnamefont {Wun}},
		\bibinfo {author} {\bibfnamefont {Z.~J.}\ \bibnamefont {Lin}}, \ and\
		\bibinfo {author} {\bibfnamefont {Y.~H.}\ \bibnamefont {Cheng}},\ }\href
	{\doibase 10.1364/JOSAB.28.002584} {\bibfield  {journal} {\bibinfo  {journal}
			{J. Opt. Soc. Am. B}\ }\textbf {\bibinfo {volume} {28}},\ \bibinfo {pages}
		{2584} (\bibinfo {year} {2011})}\BibitemShut {NoStop}%
	\bibitem [{\citenamefont {Savoia}\ \emph {et~al.}(2013)\citenamefont {Savoia},
		\citenamefont {Castaldi},\ and\ \citenamefont {Galdi}}]{Savoia:2013on}%
	\BibitemOpen
	\bibfield  {author} {\bibinfo {author} {\bibfnamefont {S.}~\bibnamefont
			{Savoia}}, \bibinfo {author} {\bibfnamefont {G.}~\bibnamefont {Castaldi}}, \
		and\ \bibinfo {author} {\bibfnamefont {V.}~\bibnamefont {Galdi}},\ }\href
	{\doibase 10.1103/PhysRevB.87.235116} {\bibfield  {journal} {\bibinfo
			{journal} {Phys. Rev. B}\ }\textbf {\bibinfo {volume} {87}},\ \bibinfo
		{pages} {235116} (\bibinfo {year} {2013})}\BibitemShut {NoStop}%
	\bibitem [{\citenamefont {Wang}\ \emph {et~al.}(2000)\citenamefont {Wang},
		\citenamefont {Grimm},\ and\ \citenamefont {Schreiber}}]{Wang:2000ta}%
	\BibitemOpen
	\bibfield  {author} {\bibinfo {author} {\bibfnamefont {X.}~\bibnamefont
			{Wang}}, \bibinfo {author} {\bibfnamefont {U.}~\bibnamefont {Grimm}}, \ and\
		\bibinfo {author} {\bibfnamefont {M.}~\bibnamefont {Schreiber}},\ }\href
	{\doibase 10.1103/PhysRevB.62.14020} {\bibfield  {journal} {\bibinfo
			{journal} {Phys. Rev. B}\ }\textbf {\bibinfo {volume} {62}},\ \bibinfo
		{pages} {14020} (\bibinfo {year} {2000})}\BibitemShut {NoStop}%
	\bibitem [{\citenamefont {Born}\ and\ \citenamefont
		{Wolf}(1999)}]{Born:1999un}%
	\BibitemOpen
	\bibfield  {author} {\bibinfo {author} {\bibfnamefont {M.}~\bibnamefont
			{Born}}\ and\ \bibinfo {author} {\bibfnamefont {E.}~\bibnamefont {Wolf}},\
	}\href@noop {} {\emph {\bibinfo {title} {{Principles of Optics}}}},\ \bibinfo
	{edition} {7th}\ ed.\ (\bibinfo  {publisher} {Cambridge University Press},\
	\bibinfo {year} {1999})\BibitemShut {NoStop}%
	\bibitem [{\citenamefont {Lang}(1987)}]{Lang:1987la}%
	\BibitemOpen
	\bibfield  {author} {\bibinfo {author} {\bibfnamefont {S.}~\bibnamefont
			{Lang}},\ }\href@noop {} {\emph {\bibinfo {title} {Linear Algebra}}},\
	\bibinfo {edition} {3rd}\ ed.\ (\bibinfo  {publisher} {Springer},\ \bibinfo
	{address} {Berlin},\ \bibinfo {year} {1987})\BibitemShut {NoStop}%
	\bibitem [{\citenamefont {Lei}\ \emph {et~al.}(2007)\citenamefont {Lei},
		\citenamefont {Chen}, \citenamefont {Nouet}, \citenamefont {Feng},
		\citenamefont {Gong},\ and\ \citenamefont {Jiang}}]{Lei:2007pb}%
	\BibitemOpen
	\bibfield  {author} {\bibinfo {author} {\bibfnamefont {H.}~\bibnamefont
			{Lei}}, \bibinfo {author} {\bibfnamefont {J.}~\bibnamefont {Chen}}, \bibinfo
		{author} {\bibfnamefont {G.}~\bibnamefont {Nouet}}, \bibinfo {author}
		{\bibfnamefont {S.}~\bibnamefont {Feng}}, \bibinfo {author} {\bibfnamefont
			{Q.}~\bibnamefont {Gong}}, \ and\ \bibinfo {author} {\bibfnamefont
			{X.}~\bibnamefont {Jiang}},\ }\href {\doibase 10.1103/PhysRevB.75.205109}
	{\bibfield  {journal} {\bibinfo  {journal} {Phys. Rev. B}\ }\textbf {\bibinfo
			{volume} {75}},\ \bibinfo {pages} {205109} (\bibinfo {year}
		{2007})}\BibitemShut {NoStop}%
	\bibitem [{\citenamefont {Kol\'a\ifmmode~\check{r}\else \v{r}\fi{}}\ and\
		\citenamefont {Ali}(1990)}]{Kolar:1990ma}%
	\BibitemOpen
	\bibfield  {author} {\bibinfo {author} {\bibfnamefont {M.}~\bibnamefont
			{Kol\'a\ifmmode~\check{r}\else \v{r}\fi{}}}\ and\ \bibinfo {author}
		{\bibfnamefont {M.~K.}\ \bibnamefont {Ali}},\ }\href {\doibase
		10.1103/PhysRevA.42.7112} {\bibfield  {journal} {\bibinfo  {journal} {Phys.
				Rev. A}\ }\textbf {\bibinfo {volume} {42}},\ \bibinfo {pages} {7112}
		(\bibinfo {year} {1990})}\BibitemShut {NoStop}%
	\bibitem [{\citenamefont {Kol\'a\ifmmode~\check{r}\else \v{r}\fi{}}\ and\
		\citenamefont {Nori}(1990)}]{Kolar:1990tm}%
	\BibitemOpen
	\bibfield  {author} {\bibinfo {author} {\bibfnamefont {M.}~\bibnamefont
			{Kol\'a\ifmmode~\check{r}\else \v{r}\fi{}}}\ and\ \bibinfo {author}
		{\bibfnamefont {F.}~\bibnamefont {Nori}},\ }\href {\doibase
		10.1103/PhysRevB.42.1062} {\bibfield  {journal} {\bibinfo  {journal} {Phys.
				Rev. B}\ }\textbf {\bibinfo {volume} {42}},\ \bibinfo {pages} {1062}
		(\bibinfo {year} {1990})}\BibitemShut {NoStop}%
	\bibitem [{\citenamefont {Vardeny}\ \emph {et~al.}(2013)\citenamefont
		{Vardeny}, \citenamefont {Nahata},\ and\ \citenamefont
		{Agrawal}}]{Vardeny:2013hj}%
	\BibitemOpen
	\bibfield  {author} {\bibinfo {author} {\bibfnamefont {Z.~V.}\ \bibnamefont
			{Vardeny}}, \bibinfo {author} {\bibfnamefont {A.}~\bibnamefont {Nahata}}, \
		and\ \bibinfo {author} {\bibfnamefont {A.}~\bibnamefont {Agrawal}},\
	}\href@noop {} {\bibfield  {journal} {\bibinfo  {journal} {Nat. Photon.}\
		}\textbf {\bibinfo {volume} {7}},\ \bibinfo {pages} {177} (\bibinfo {year}
		{2013})}\BibitemShut {NoStop}%
\end{thebibliography}


\newpage

%############################################################
%                Figure1
%
%############################################################

%############################################################
%                Figure2
%
%############################################################

%############################################################
%                Figure3
%
%############################################################

%############################################################
%                Figure4
%
%############################################################

%############################################################
%                Figure5
%
%############################################################

%############################################################
%                Figure6
%
%############################################################

%############################################################
%                Figure7
%
%############################################################

%############################################################
%                Figure8
%
%############################################################

%############################################################
%                Figure9
%
%############################################################

%############################################################
%                Figure10
%
%############################################################

%############################################################
%                Figure11
%
%############################################################

%############################################################
%                Figure12
%
%############################################################

%############################################################
%                Figure13
%
%############################################################

\end{document}